\documentclass [aps,pra,amssymb,amsmath,twocolumn,showpacs]{revtex4-1}

\usepackage{amsmath}
\usepackage{amssymb}

\usepackage{graphicx}
\usepackage{verbatim}
\usepackage{bm}
\usepackage{color}
\usepackage{graphicx,epsfig,amsfonts,amssymb}
\usepackage{bm}
\usepackage{times}

\usepackage[all]{xy}

\newcommand{\nn}{\nonumber}

\newcommand{\ra}{\rangle}
\newcommand{\rar}{\rightarrow}

\newcommand{\ua}{\uparrow}
\newcommand{\be}{\begin{eqnarray}}
\newcommand{\ee}{\end{eqnarray}}

\newcommand{\bs}{\begin{equation}\begin{split}}
\newcommand{\es}{\end{split}\end{equation}}

\date{\today}

\begin{document}
\title{Multitime  Landau-Zener model: classification of solvable Hamiltonians}

\author{Vladimir Y. Chernyak$^{a, b}$, Nikolai A. Sinitsyn$^c$, Chen Sun$^d$}
\affiliation{$^a$Department of Chemistry, Wayne State University, 5101 Cass Ave, Detroit, Michigan 48202, USA}
\affiliation{$^b$Department of Mathematics, Wayne State University, 656 W. Kirby, Detroit, Michigan 48202, USA}
\affiliation{$^c$Theoretical Division, Los Alamos National Laboratory, Los Alamos, NM 87545, USA}
\affiliation{$^d$Department of Physics, Brown University, Providence, Rhode Island 02912, USA}

\begin{abstract}
We discuss a class of models that generalize the two-state Landau-Zener (LZ) Hamiltonian to both the multistate and multitime evolution. It is already known that the corresponding quantum mechanical evolution can be understood in great detail. Here, we present an approach to classify such solvable models, namely, to identify all their independent families for a given number $N$ of interacting states and prove the absence of such families for some types of interactions. We also discuss how, within a solvable family, one can classify the scattering matrices, i.e., the system's dynamics. Due to the possibility of such a detailed classification,  the multitime Landau-Zener (MTLZ) model defines a useful special function of theoretical physics.   
\end{abstract}

\maketitle

\section{Introduction}
\label{sec:intro}

Modern quantum science encounters considerable new mathematical challenges. In studies of explicitly time-dependent phenomena, such as quantum annealing, dynamic phase transitions, control of quantum devices, quenching, and thermalization  we deal with strongly nonequilibrium complex  systems that are often not accessible to numerical modeling at the desired scales.  Standard  analytical tools, however, are too limited for such applications. Efforts to keep physical models analytically tractable often lead to considerable over-simplifications, so that generalizations become  unjustified and misleading.
It is desirable to develop tools to study explicitly time-dependent Hamiltonians with complexity of, e.g., models that are commonly studied by  methods of the conformal field theory and the Bethe ansatz.

Recently \cite{commute}, the conditions on a time-dependent Hamiltonian were found that lead to considerable understanding of the corresponding dynamics. However, the integrability conditions in \cite{commute} only provided a test for integrability.
So far, there is no straightforward path to a systematic classification of such integrable models. Integrability conditions have been used either to validate hypotheses or to generate integrable time-dependent models within the already known class of Gaudin magnet Hamiltonians \cite{yuzbashyan}.

In this article, we develop an approach for detailed classification of solvable time-dependent Hamiltonians that have a specific unifying property.
We will also discuss that there is an analytical solution for the corresponding scattering problem. Hence, this solvable family defines  an unusual new special function that can play a similar role in  complex time-dependent quantum physics as the parabolic cylinder function plays in  time-dependent two-state physics \cite{landau,zener,majorana,stuckelberg}. 


This article is organized as follows. Sections II and III are still introductory. In II, we define the class of models that we will mainly consider -- the multitime Landau-Zener (MTLZ) models, and in III we discuss the difference between separable and nontrivial integrable models. In section IV, we show that integrability conditions for MTLZ models can be conveniently presented by the data on graphs, and sketch a scheme for retrieving independent solvable models for such data. Sections V to VIII are applications of the graph method described in section IV to specific graph geometries with $N \le 10$.  Discussions and conclusions are  left to section IX.

\section{Linearly time-dependent Hamiltonians}
The simplest time-dependent  Hamiltonian is the one with linearly changing parameters. Hence, we will consider the Schr\"odinger equation
\begin{equation}
i\frac{d}{dt}\psi={H}(t)\psi,\quad  {H}(t) = {A} +{B}t,
\label{multistate LZ}
\end{equation}
where   ${A}$ and ${B}$ are constant Hermitian $N\times N$ matrices with real entries (we set $\hbar=1$), and $\psi$ is a vector with $N$ components.
It is a natural generalization of the two-state Landau-Zener (LZ) model for spin 1/2 in a linearly time-dependent magnetic field \cite{zener,majorana}. The eigenvalues  and eigenvectors of the matrix $Bt$ are called diabatic energies and diabatic states, respectively.  Off-diagonal elements of the matrix $A$, in the basis of diabatic states, are called couplings.  Any two diabatic states are called coupled if the matrix element of $A$ between these states is nonzero.   

The general solution of Eq.~(\ref{multistate LZ}) is not known. Nevertheless,
in our recent article \cite{large-class}, we pointed that considerable understanding can be obtained when such Hamiltonians create a family
 of  some $M>1$ Hamiltonians of the form
\begin{eqnarray}
\label{linear-family} H_{j} (\bm{x}) = B_{kj} x^{k} + A_{j}, \quad j,k=1,\ldots, M,
\end{eqnarray}
where $ {\bm{x}} =(x^1,\ldots, x^M)$ is called a time-vector, and $B_j$, $A_j$ are real symmetric matrices; here and in what follows we will assume summations over repeated upper and lower indices.
Within this family, the state vector must satisfy simultaneously $M$ equations:
\begin{equation}
\label{system1}
 i \partial \psi(\bm{x})/\partial  x^j  = {H}_{j} (\bm{x}) \psi(\bm{x}), \; \phantom{\sum} j =  1, \ldots, M, \quad M>1.
\end{equation}
The parameter $x^j$ in $H_{j}$ can be identified with the physical time.
Note that if we set $x^k=\rm{const}$  in (\ref{system1}) for all $k\ne j$ and identify $x^j$ with $t$, then each of the equations (\ref{system1}) becomes an  independent multistate LZ  model of the form (\ref{multistate LZ}). Moreover, the evolution of the system (\ref{system1}) along a path given by a linear combination of time variables $x^{j}$ is equivalent to a model of the form (\ref{multistate LZ}).
For this property, the system of equations (\ref{system1}) with the set of Hamiltonians of the form (\ref{linear-family}) was called the {\it multitime Landau-Zener} (MTLZ) model.

According to  Ref.~\cite{commute},  an MTLZ  system may be solvable if equations (\ref{system1}) are consistent with each other. For real symmetric matrices $H_j$ this happens when two conditions are satisfied:
\be
[{H}_i,{H}_j]=0,
\label{cond1}
\ee
\be
\partial {H}_i/\partial x^j =  \partial {H}_j/\partial x^i,  \quad i,j=1,\ldots, M.
\label{cond2}
\ee
We will call \eqref{cond1} and  \eqref{cond2} the integrability conditions.

In Ref.~\cite{large-class},  we focused only on the general properties of MTLZ systems. In particular, we already proved that the scattering problem for any multistate LZ model that can be generated from such a family can be solved explicitly in terms of the matrix product ansatz, and that parameters of such models are constrained to have several common properties. For example, we showed that,  when plotted as functions of one time variable  $x^j$, the energy levels of the Hamiltonian $H_j$ from the family (\ref{linear-family})  pass through a known number of exact crossing points. Here, in contrast, we are going to discuss classification of such systems. 

\section{Separable and nontrivial integrable models}
A trivial example of  an integrable family (\ref{linear-family}) is found among Hamiltonians of noninteracting spins that experience independent LZ evolution \cite{large-class}:
\begin{eqnarray}
\label{nonint1}
&&\nonumber H(t) ={\cal H}_1 \otimes \hat{1}_2 \otimes \cdots \otimes  \hat{1}_2 +\hat{1}_2 \otimes {\cal H}_2 \otimes \hat{1}_2 \otimes \cdots\otimes  \hat{1}_2+\\
&& \ldots+   \hat{1}_2 \otimes \cdots \otimes \hat{1}_2 \otimes {\cal H}_N,
\ee
where $\hat{1}_2$ is a unit matrix acting in the space of the corresponding spin and
\be
{\cal H}_k= \left(\begin{array}{cc}
 \beta_{1k} t +\epsilon_{1k} & \gamma_k \\
\gamma_k &  \beta_{2k} t +\epsilon_{2k}
\end{array} \right), \quad k=1,\ldots, N,
\label{hlz-1}
\ee
are the two-state LZ Hamiltonians with different constant parameters $\beta_{1,2}$, $\epsilon_{1,2}$ and $\gamma$. Any such $H(t)$ has $N-1$ linearly independent Hamiltonians with the same structure and satisfying relations (\ref{cond1})-(\ref{cond2}).

The solution of the Schr\"odinger equation for the model (\ref{hlz-1}) is also trivial. Since all spins are independent, the evolution operator is a direct product of such operators for each spin:
\be
U(t) =U_1(t) \otimes \cdots \otimes U_N(t),
\label{evop}
\ee
where any $U_k(t)$ is known because the LZ model is solved in terms of the parabolic cylinder functions.

Despite the simplicity of the model (\ref{nonint1}), many studies of quantum annealing and a dynamic passage through a phase transition were based on reducing a problem to independent two-state dynamics. A notable example is the Ising spin chain in a transverse magnetic field
\cite{kzt1,kzt3}.
Our goal, however, is to find MTLZ families that do not have the trivial form (\ref{nonint1}) in any fixed basis.
An example of a nontrivial family is the $\gamma$-magnets \cite{gammamagnet}:
\be
H_1(t,\varepsilon) =\varepsilon \prod_{j=1}^N \sigma_{j}^{z}  +\sum_{j=1}^N  \left( \beta_j t \sigma_{j}^{z} + g_j {\gamma_j} \right),
\label{ham-string1}
\ee
\be
H_2 (t,\varepsilon) =t \prod_{j=1}^N \sigma_{j}^{z} +\sum_{j=1}^N  \left(\frac{ \varepsilon}{ \beta_j}  \sigma_{j}^{z} + \frac{g_j}{\beta_j} \tilde{\gamma_j} \right).
\label{ham-string2}
\ee
where  $\beta_i$, and $g_i$ are constant parameters; $\sigma^{\alpha}_j$ are the Pauli operators for $j$th spin, and
\begin{eqnarray}
\label{string1}
\nonumber && \gamma_1\!=\! \sigma_{1}^{x}, \,\,\, \gamma_2 \!=\! \sigma_{2}^{x} \sigma_{1}^{z}, \,\,\,\cdots, \,\,\, \gamma_N \!=\! \sigma_{N}^{x} \prod \limits_{k=1}^{N-1} \sigma_{k}^{z}, \\
\label{string2}
 && \tilde{\gamma}_1\!=\! \sigma_{1}^{x} \prod \limits_{k=2}^{N} \sigma_{k}^{z} ,\,\,\,\cdots,\,\,\,  \tilde{\gamma}_{N-1} \!=\!\sigma^{x}_{N-1} \sigma^{z}_{N},  \,\,\, \tilde{\gamma}_N \!=\!\sigma^{x}_{N}. 
\end{eqnarray}
For the two-time vector
\be
{\bm \tau} \equiv (t, \varepsilon),
\label{twotime}
\ee
$H_1$ and $H_2$ satisfy (\ref{cond1}) and (\ref{cond2}).

In Fig.~\ref{spectrum-16}, we plot the energy spectrum of the Hamiltonian $H_1(t)$ for different values of $t$ and fixed other parameters. Integrability of this model can be inferred from the large number of points with  exact crossings of energy levels.
According to \cite{large-class}, the number of such exact pairwise level crossings in solvable multistate LZ models should be the same as the number of zero direct couplings between the diabatic states. This property holds true for the Hamiltonian $H_1$.
\begin{figure}[!htb]
\scalebox{0.25}[0.25]{\includegraphics{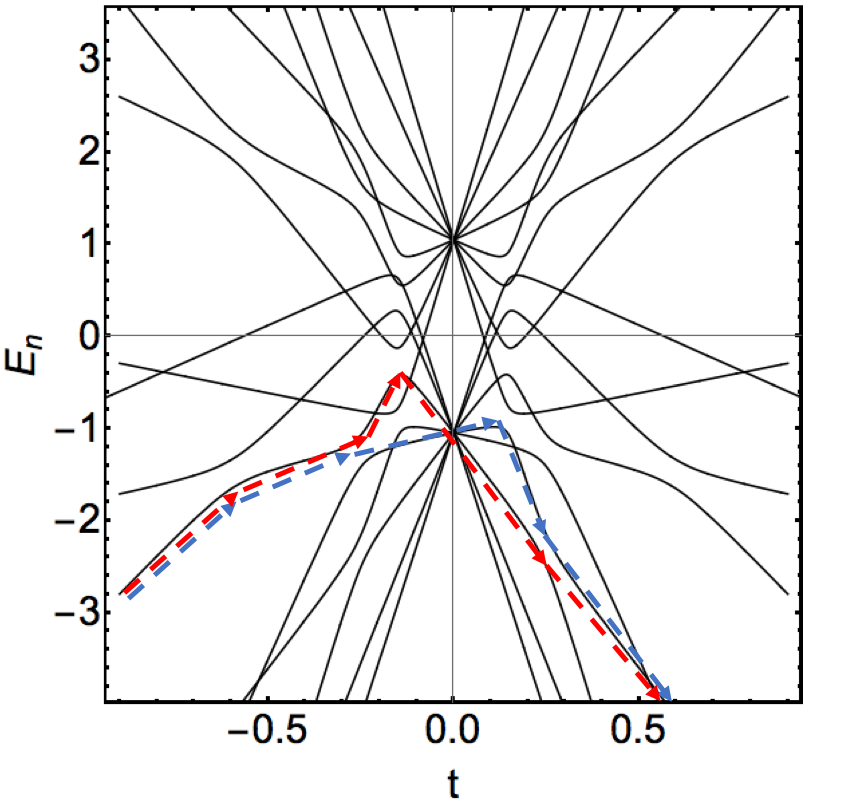}}
\caption{Eigenvalues of the Hamiltonian (\ref{ham-string1}) as function of $t$ for $N=4$ interacting spins. If we count a simultaneous exact crossing of $n$ levels in one point as $n(n-1)/2$ pairwise exact level crossings, then this figure contains 88 exact pairwise level
crossings, as it is required by integrability conditions in the multistate LZ theory. The blue and red arrows show an example of interfering semiclassical trajectories. The choice of the parameters: $e=1$, $\beta_1=0.5$, $\beta_2=1.7$, $\beta_3=4.1$, $\beta_4=7.1$, $g_1=0.14$, $g_2=0.15$, $g_3=0.17$, $g_4=0.15$.}
\label{spectrum-16}
\end{figure}

The diabatic states are the eigenstates of the time-dependent part of the Hamiltonian. In the model (\ref{ham-string1}), they are the spin projection states along the $z$ axis, such as $|\ua \ua\ldots \ua \ra$.
 According to the adiabatic theorem, when energy levels are well separated, transitions between them are suppressed. This happens for the spectrum in Fig.~\ref{spectrum-16} as $t\rar \pm \infty$. However, for the time interval and the parameters that we used in this figure, different pairs of levels experience avoided crossings, i.e. places where levels do not cross exactly but appear very close to each other for a short time interval.
After passing them, the system has finite amplitudes to stay on the initial level or to jump to a new one.  Thus, for evolution from $t=-\infty$ to $t=+\infty$, one can estimate the amplitude of transitions between any pair of diabatic states by drawing all semiclassical trajectories that connect the initial state at $t=-\infty$ and the final state at $t=+\infty$, and then summing the amplitudes of these trajectories for a given transition.

 A common feature of all $\gamma$-magnets with $N>1$ is that there are generally more than one trajectory connecting an arbitrary pair of energy levels that correspond to some pair of the Hamiltonian eigenstates at $t=\pm \infty$.  
  An example is shown by red and blue arrows in Fig.~\ref{spectrum-16}.
 This property, rather than the presence of exact crossings, is the signature of purely quantum and nontrivial behavior.  For example, consider the separable Hamiltonian for four spins:
 \be
 H_{sep}=\sum_{i=1}^4 \left[ (\beta_it +\epsilon_i)\sigma_z^i + g_i \sigma_x^i \right].
 \label{noint2}
 \ee
 The spectrum for $H_{sep}$ is shown in Fig.~\ref{separable-fig}. It contains the same number, 88, of the exact pairwise crossing points, as Fig.~\ref{spectrum-16}. However, careful examination of Fig.~\ref{separable-fig} shows that in the semiclassical picture there is always only a single trajectory that connects one level at $t=-\infty$ with another level at $t=+\infty$.

\begin{figure}[!htb]
\scalebox{0.12}[0.12]{\includegraphics{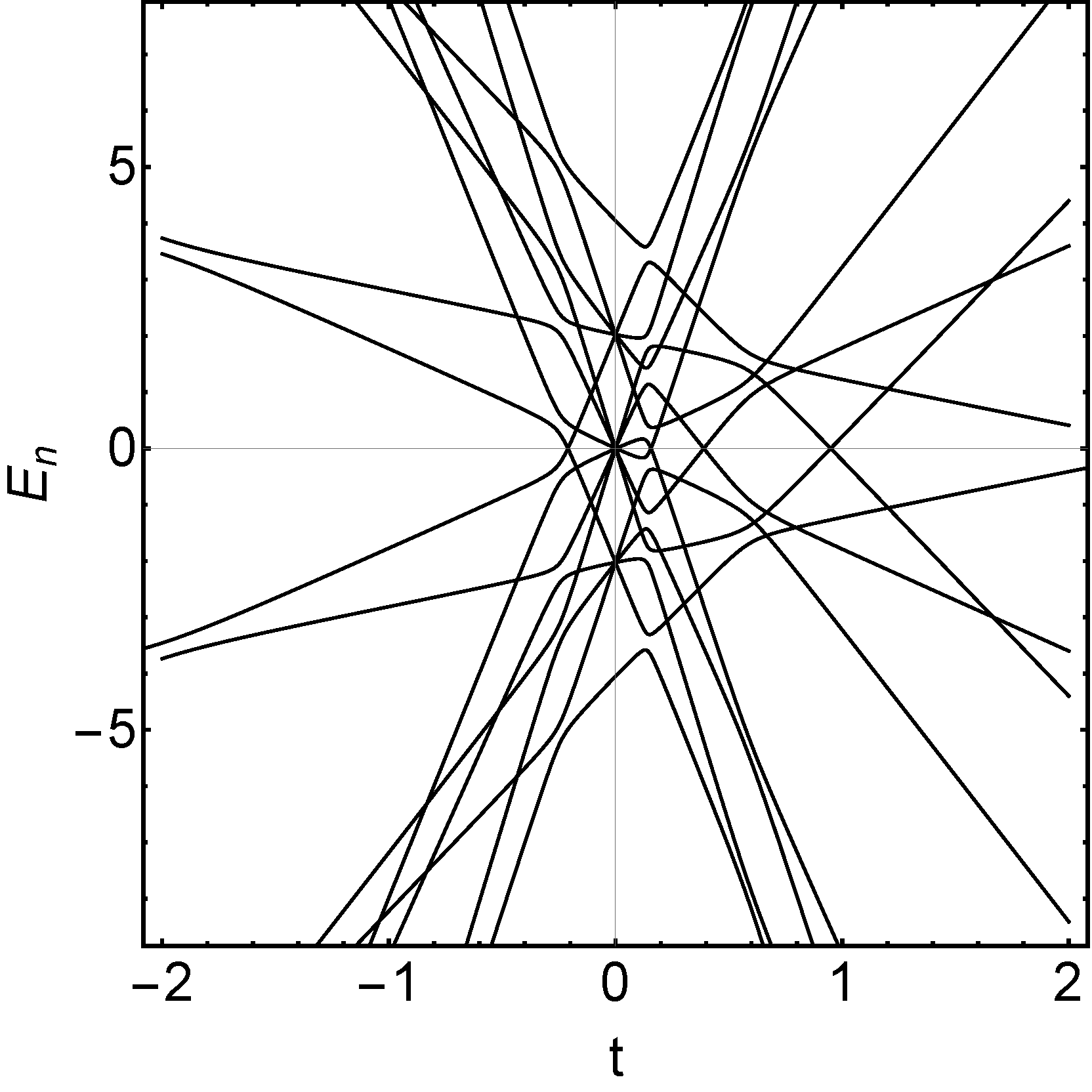}}
\caption{Time-dependent spectrum of  the separable Hamiltonian  (\ref{noint2}) as a function of $t$ for $N=4$ interacting spins. Here $\epsilon_i=(-1)^i \varepsilon$, and all other parameters are as in Fig.~\ref{spectrum-16}.}
\label{separable-fig}
\end{figure}

 \begin{figure}[!htb]
\scalebox{0.32}[0.32]{\includegraphics{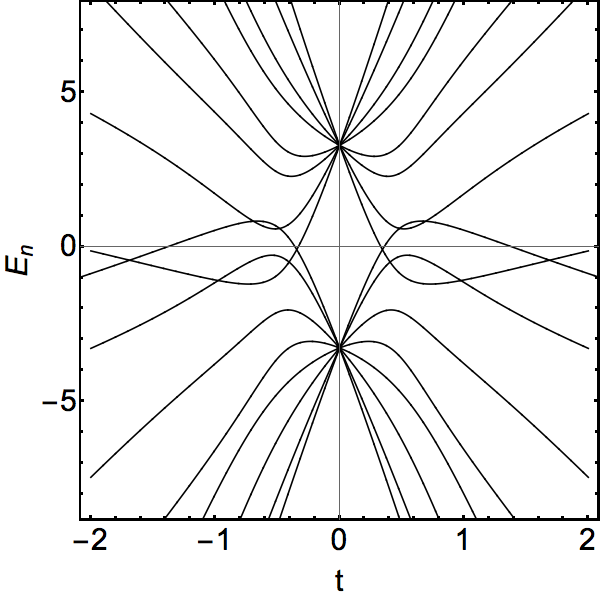}}
\caption{Spectrum of the same $\gamma$-magnet model as in Fig.~\ref{spectrum-16} but for ten times larger couplings $g_i$, $i=1,2,3,4$. Several pairs of crossings annihilated each other in comparison to  Fig.~\ref{spectrum-16}.}
\label{spectrum-fig2}
\end{figure}
 Another feature of  nontrivial interactions in the $\gamma$-magnet is that the number of exact crossings is not conserved at large values of off-diagonal couplings.
 The theory in \cite{large-class} guarantees 88 such crossings  for $H_1$ at $N=4$  for finite but only sufficiently small values of $g_i$. For example, if we increase all couplings of the model in Fig.~\ref{spectrum-16} ten times, we find the spectrum shown in Fig.~\ref{spectrum-fig2} with fewer exact pairwise crossings. Such a reduction does not happen with the spectrum of the separable spin model (\ref{noint2}) because exact crossings there are guaranteed by the lack of spin-spin interactions.
Thus, both the spectrum and semiclassical analysis of $\gamma$-magnets show features that are not present in the separable spin models.

This comparison between separable and non-separable integrable models suggests that the latter may describe considerably more complex dynamics. In what follows, we will develop an approach to classify all MTLZ models on the same footings. We will find, for example, that 
the separable model corresponds only to a very special and very symmetric case in such a classification, whereas the Hamiltonian of the general case is described by a considerably bigger set of parameters.

\section{Integrability conditions for MTLZ families on 
graphs}
\label{sec:integr-cond-LF-LA}

Originally, we constructed the 
family (\ref{ham-string1})-(\ref{ham-string2}) using the trial-and-error approach. More systematic classification of such solvable families is needed. 
Thus, we want to know whether there are restrictions on the numbers of independent Hamiltonians in such families, or whether we can add extra nontrivial interaction terms in such Hamiltonians without breaking integrability. 

Substituting Eq.~\eqref{linear-family} into (\ref{cond1}) and (\ref{cond2}) we find matrix relations  for a MTLZ family \cite{large-class}:
\begin{eqnarray}
\label{linear-family-2} && B_{kj} = B_{jk}, \;\;\; [B_{jk}, B_{lm}] = 0, \\
\label{linear-family-3} && [B_{sj}, A_{k}] - [B_{sk}, A_{j}] = 0, \\
\label{linear-family-4} &&[A_{j}, A_{k}] = 0,  \qquad k,j,l,m,s=1,\ldots M.
\end{eqnarray}

Note that here the lower indices  are not indices of matrix elements but rather indices that enumerate independent Hamiltonians in an MTLZ family. We will call the number of independent Hamiltonians, $M$, the {\it dimension of the MTLZ family}.  Equations~(\ref{linear-family-2})-(\ref{linear-family-4}) are the integrability conditions for MTLZ models. Due to Eq.~(\ref{linear-family-2}), all matrices $B_{jk}$ can be diagonalized in the same orthonormal basis set $(\bar{\bm{e}}_{a} \, | \, a = 1, \ldots, N)$, namely, the set of states that we will call  the {\it diabatic states}.

\subsection{Algebra of forms on the connectivity graph}
\label{sec:integr-cond-LF-LA-graphs}

In order to satisfy conditions (\ref{linear-family-2})-(\ref{linear-family-4}), the real symmetric matrices $A_j$  must have some zero matrix elements in the diabatic basis.
Therefore, with any MTLZ family of models, it is convenient to associate an undirected graph $\Gamma = (\Gamma_{0}, \Gamma_{1})$, whose vertices $a \in \Gamma_{0}$ ($a=1,\ldots, N$) represent the diabatic basis states and  edges $\alpha \in \Gamma_1$ correspond to the nonzero couplings between the diabatic states. We will call $\Gamma$ the connectivity graph.
In what follows, it will also be useful to assume that edges have orientations, which we will mark by arrows.
For example, the family of models with the Hamiltonians of the form
\begin{align}\label{h-four-ex}
 H(t)=\left( \begin{array}{cccc}
\beta_1 t+e_1 & g_{12} & 0 & g_{14} \\
g_{12} & \beta_2 t+e_2 &  g_{23} &  0 \\
0 & g_{23} &  \beta_3 t+e_3  & g_{34} \\
g_{14}  & 0 & g_{34}  & \beta_4 t+e_4  \\
\end{array}\right),
\end{align}
where $g_{ij}$, $\beta_i$ are some constant parameters, has the connectivity graph shown in Fig.~\ref{4-loop-fig}. The meaning of  its edge orientations will be explained later.
\begin{figure}[!htb]
\scalebox{0.3}[0.3]{\includegraphics{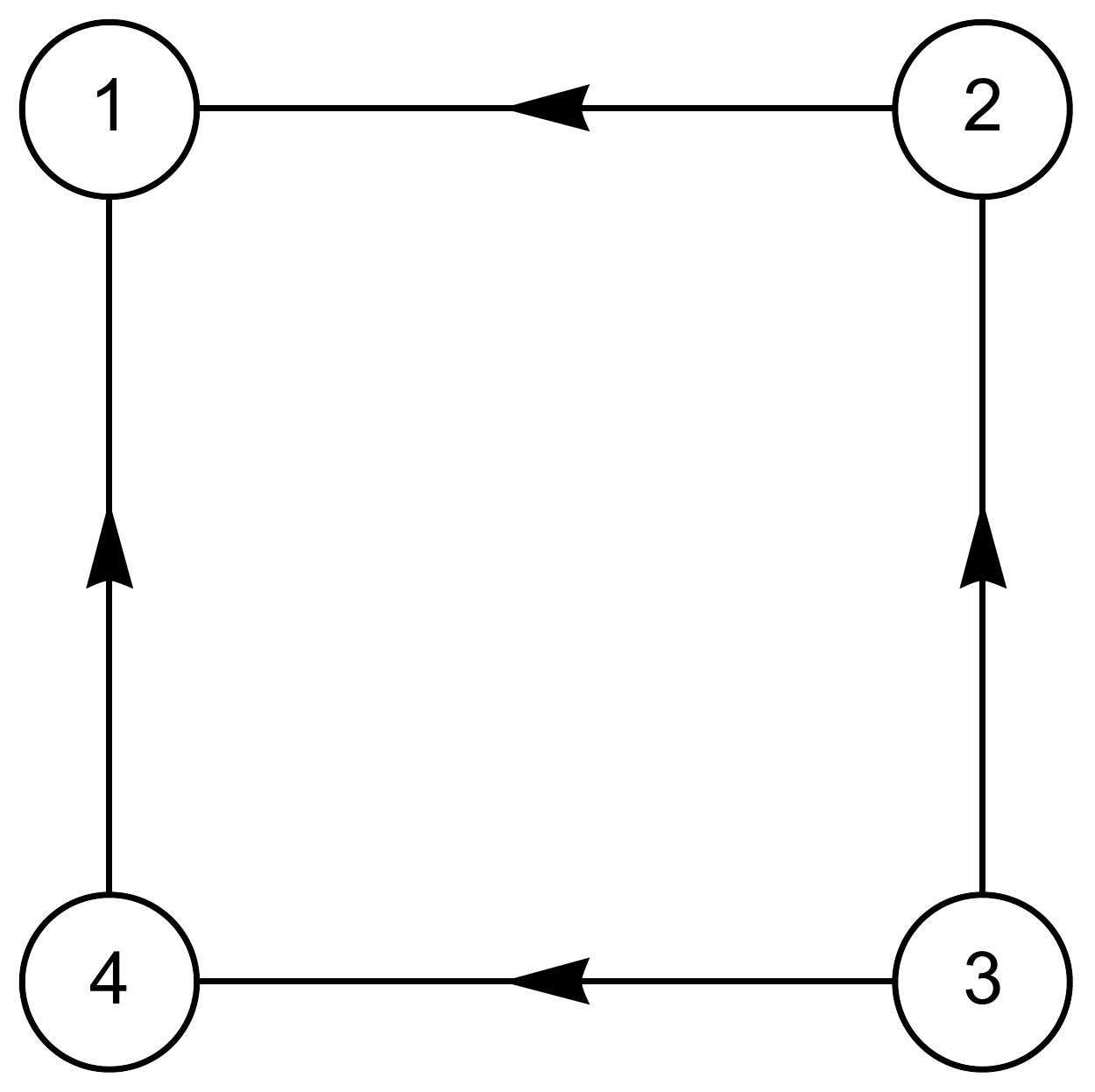}}
\caption{Directed graph representing a four-state model (\ref{h-four-ex}).}
\label{4-loop-fig}
\end{figure}

Let $\Lambda_{kj}^a$, $a=1,\ldots, N$ be eigenvalues of the matrices $B_{kj}$.
Since  $B_{kj} = B_{jk}$, due to Eq.~(\ref{linear-family-2}), we have the obvious symmetric property $\Lambda_{jk}^{a} = \Lambda_{kj}^{a}$. Hence, each vertex $a$ is the residence for a quadratic (symmetric bilinear) form
\begin{eqnarray}
\label{Lambda-quadr-form} \Lambda^{a} = \Lambda_{jk}^{a} dx^{j} \otimes dx^{k},
\end{eqnarray}
where ``$\otimes$'' denotes the tensor direct product. The nonzero couplings $A_{j}^{ab}$ will be naturally considered as $j$-components of a 
linear form
\begin{eqnarray}
\label{A-1-form}
A^{ab} = A^{ba} = A_{j}^{ab} dx^{j}.
\end{eqnarray}
Note that this notation  resolves the first constraint, given by Eq.~(\ref{linear-family-2}), automatically because the symmetry of the 2-form in (\ref{Lambda-quadr-form}) means that $ \Lambda_{jk}= \Lambda_{kj}$.

The second constraint (Eq.~(\ref{linear-family-3})), now has a form
\begin{eqnarray}
\label{constr-2-LA} (\Lambda^{a} - \Lambda^{b}) \wedge A^{ab} = 0, \;\;\; \forall \{a, b\} \in \Gamma_{1}, 
\end{eqnarray}
where ``$\wedge$'' denotes the skew symmetric tensor product (the wedge product).
Equation~(\ref{constr-2-LA}) is straightforward to verify by substituting (\ref{Lambda-quadr-form}) and (\ref{A-1-form}) into the left hand side of (\ref{constr-2-LA}) and compare the antisymmetrized over $j$ and $k$ coefficients near $(dx^j \wedge dx^k) \otimes dx^s$ with (\ref{linear-family-3}).
Since the vanishing of the wedge product of two vectors is equivalent to their linear dependence, Eq.~(\ref{constr-2-LA}) is equivalent to
\begin{eqnarray}
\label{constr-2-LA-2} \Lambda^{a} - \Lambda^{b} = \chi^{ab} \otimes A^{ab},
\end{eqnarray}
for some linear nonzero form $\chi^{ab} = - \chi^{ba}$. Due to the property $\Lambda_{ij}^a=\Lambda_{ji}^a$,  Eq.~(\ref{constr-2-LA-2}) implies that $\chi_i^{ab}A_j^{ab}= \chi_j^{ab} A_i^{ab}$, or in other words: $\chi^{ab} \wedge A^{ab} = 0$. This  implies the linear dependence of $\chi^{ab}$ and $A^{ab}$, which being substituted into Eq.~(\ref{constr-2-LA-2}) results in
\begin{eqnarray}
\label{constr-2-LA-3} \gamma^{ab} (\Lambda^{a} - \Lambda^{b}) = A^{ab} \otimes A^{ab},
\end{eqnarray}
for some $\gamma^{ab} = -\gamma^{ba} \ne 0$. 
Using the introduced notation the third constraint (Eq.~(\ref{linear-family-4})) can be naturally represented as
\begin{eqnarray}
\label{constr-3-LA}
 \sum_{s \in {\cal P}_{2}(a, b)} (A_{j}^{s_{2}} A_{k}^{s_{1}} - A_{k}^{s_{2}} A_{j}^{s_{1}}) = 0 \;\;\; \forall \, a, b \in \Gamma_{0},
\end{eqnarray}
where the summation goes over all length $2$ paths on the graph, and we denoted by ${\cal P}_{l}(a, b)$ the set of paths $s = (s_{1}, \ldots, s_{l})$, with $s_{j} \in \Gamma_{1}$ for $j = 1, \ldots, l$, that starts at $a$ and ends at $b$.
Equations~(\ref{constr-2-LA-3})-(\ref{constr-3-LA}) are homogeneous of degree $2$. They are equivalent to the integrability conditions (\ref{linear-family-2})-(\ref{linear-family-4}) but they are simpler for analysis for a given connectivity graph.



Finally, we note that in multistate LZ theory it is assumed that the directly coupled diabatic energy levels must cross. For levels $a$ and $b$, this happens on the hypersurface that is defined by conditions
\be
\left( \Lambda_{ij}^a \! -\! \Lambda_{ij}^b  \right) x^j \! = \!0, \,\,\, a,b \! =\!1,\ldots, N, \,\,\, i\!=\!1,\ldots, M.
\label{cross-cond1}
\ee
Using (\ref{constr-2-LA-3}), we find that this condition can be rewritten in terms of $A^{ab}_{i}A^{ab}_j$.   Suppose now that there is a diabatic state $c$ such that $A^{ab}$ and $A^{ac}$ are linearly dependent. We find then that if condition (\ref{cross-cond1}) is satisfied for levels $a$ and $b$, it is also satisfied for $a$ and $c$. In other words, levels $a$, $b$, and $c$ cross simultaneously.

The multistate LZ models with simultaneous multiple diabatic level crossings are widely known and used in practice (see e.g., Ref.~\cite{cross} that is fully devoted to them). However,  all of them are likely derivable as limits of models with only pairwise level intersections. Therefore, in this article we will restrict our studies only to the Hamiltonians without triple or higher order intersections, in one point, of directly coupled diabatic levels. In the graph language, this means that $A^{\alpha}$ forms have the following property:  for any pair of distinct edges $\alpha, \beta \in \Gamma^{1}$ that share a vertex, that is $\alpha \cap \beta \ne \emptyset$, the forms $A^{\alpha}$ and $A^{\beta}$ are linearly independent. We will call a family that satisfies this property a \underline{\it good family}.
Let us now show that restricting our studies to the good families leads to considerable additional simplifications.

\subsection{Refined integrability conditions for MTLZ models}\label{sec:properties}
Let us define  a cyclic path
\be
n = \sum_{\alpha \in \Gamma_{1}} n_{\alpha} \alpha
\label{n-graph}
\ee
on the graph $\Gamma$,  as a  combination of the edges $\alpha$ on $\Gamma$ with the zero boundary; here $n_{\alpha}=\pm$ account for possible orientations of the edges.
Namely, let us define $s^{ab} =  {\rm sgn}(\gamma^{ab}) = \pm 1$, for $\alpha = \{a, b\}$. We can then represent $\gamma^{ab} = s^{ab} \gamma^{\alpha}$, 
where  $\gamma^{\alpha} = |\gamma^{ab}|$. For any cyclic path $n$,  we can now prescribe the coefficients $n_{\alpha} =\pm 1$ to all its edges:  $n_{\alpha}=s^{ab}$ if $n$ passes the edge from $a$ to $b$.
Hence, signs of $\gamma^{ab}$ define unique directions of edges along any loop of the graph.

A vertex $a$ will be called a {\it source} or a {\it sink} if $s^{ab} = -1$, or $s^{ab} = 1$, respectively, for all $\{a, b\} \in \Gamma_{1}$, i.e., if all arrows point, respectively, out or  in. 
A vertex will be called {\it intermediate} if it is not  a sink and  not a source for all edges. For example, the arrows on the edges in Fig.~\ref{4-loop-fig} mean that vertex 1 is a sink, vertex 3 is a source, and vertices 2 and 4 are intermediate, and the signs are $s^{12}=s^{14}  =-s^{32}=-s^{34}= 1$. We will call  $\gamma^{\alpha}$ the LZ parameters for their similarity with the analogous combination that enters the transition amplitude in the simple two-state LZ formula \cite{zener}.

It is now convenient to introduce the rescaled forms
\begin{eqnarray}
\label{define-bar-A} \bar{A}^{\alpha} = \frac{A^{ab}}{\sqrt{|\gamma^{ab}|}} = \frac{A^{\alpha}}{\sqrt{\gamma^{\alpha}}}.
\end{eqnarray}
For any cycle $n = \sum_{\alpha \in \Gamma_{1}} n_{\alpha} \alpha$, the integrability conditions can then be written as
\begin{eqnarray}
\label{cycle-property-bar-A} \sum_{\alpha \in n} n_{\alpha} \bar{A}^{\alpha} \otimes \bar{A}^{\alpha} = 0,
\end{eqnarray}
 and $ \forall \, a, b \in \Gamma_{0}$:
\begin{eqnarray}
\label{constr-3-LA-bar-A} \sum_{c \in \Gamma_{0}}^{\{a, c\}, \{b, c\} \in \Gamma_{1}} \sqrt{\gamma^{\{a, c\}}\gamma^{\{b, c\}}} \bar{A}^{\{a, c\}} \wedge \bar{A}^{\{b, c\}} = 0.
\end{eqnarray}

From Eq.~(\ref{constr-3-LA-bar-A}) follows the following property:

(i) any pair ${\alpha, \beta}$ of edges that share a vertex belongs to at least one 
length-$4$ loop.

In appendix~\ref{sec:integr-cond-LF-LAG-properties}, we also prove two properties that strongly restrict the types of  graphs that can sustain integrable families. Namely,

(ii) the graph $\Gamma$ must not have length $3$ simple loops;

(iii) the vector space spanned by the four $\bar A$ forms on any length-$4$ loop has dimension $2$.



Now we summarize the program for how to retrieve the integrable families for a given connectivity graph. First, we should check whether  conditions (i) and (ii) are satisfied. If not, then there is no
integrable family for this graph. Otherwise, we take the following steps:

1) We start with choosing the orientations on the graph, namely, fixing the sign $s^{ab}$ on every edge $\alpha=\{a,b\}$.

2) We further identify/classify the solutions of Eq.~(\ref{cycle-property-bar-A}), viewed as a system of bilinear equations on the forms $\bar{A}^{\alpha}$; the 
number of independent equations is given by the number of independent $1$-cycles on the graph. Generally,  solution of Eq.~(\ref{cycle-property-bar-A}) is not unique but rather  depends on  free parameters, which we will call {\it rapidities}.

3) Once $\bar{A}^{\alpha}$ are identified, we find the solutions of Eq.~(\ref{constr-3-LA-bar-A}), viewed as a system of bilinear equations for $\sqrt{\gamma^{\alpha}}$; we will show later that any particular equation has a very simple and scalar form. Again, the solution may not determine all $\gamma^{\alpha}$ uniquely, so some of  $\gamma^{\alpha}$ then become free parameters of the model. At this stage, having Eq.~(\ref{define-bar-A}), we can reconstruct couplings of the Hamiltonians, which will depend on rapidities and  $\gamma^{\alpha}$.

4) Finally, the quadratic
forms $\Lambda^{a}$
associated with the vertices are  obtained with Eq.~(\ref{constr-2-LA-3}). Again, this equation may not fix all $\Lambda^{a}$. The parameters that describe this freedom also become free parameters of the MTLZ Hamiltonians.

Note that within such a scheme, the forms $\bar{A}^{\alpha}$ are obtained in some (abstract) basis set. The dimension of the vector space spanned onto $\bar{A}^{\alpha}$ is the actual dimension of the MTLZ family. Multistate LZ models within this family are related up to an invertible linear transformation in the space of free parameters.

\section{Four-vertex graph}
\label{sec:integr-cond-LF-LAG-4loop}
\begin{figure}[!htb]
(a)~ \scalebox{0.3}[0.3]{\includegraphics{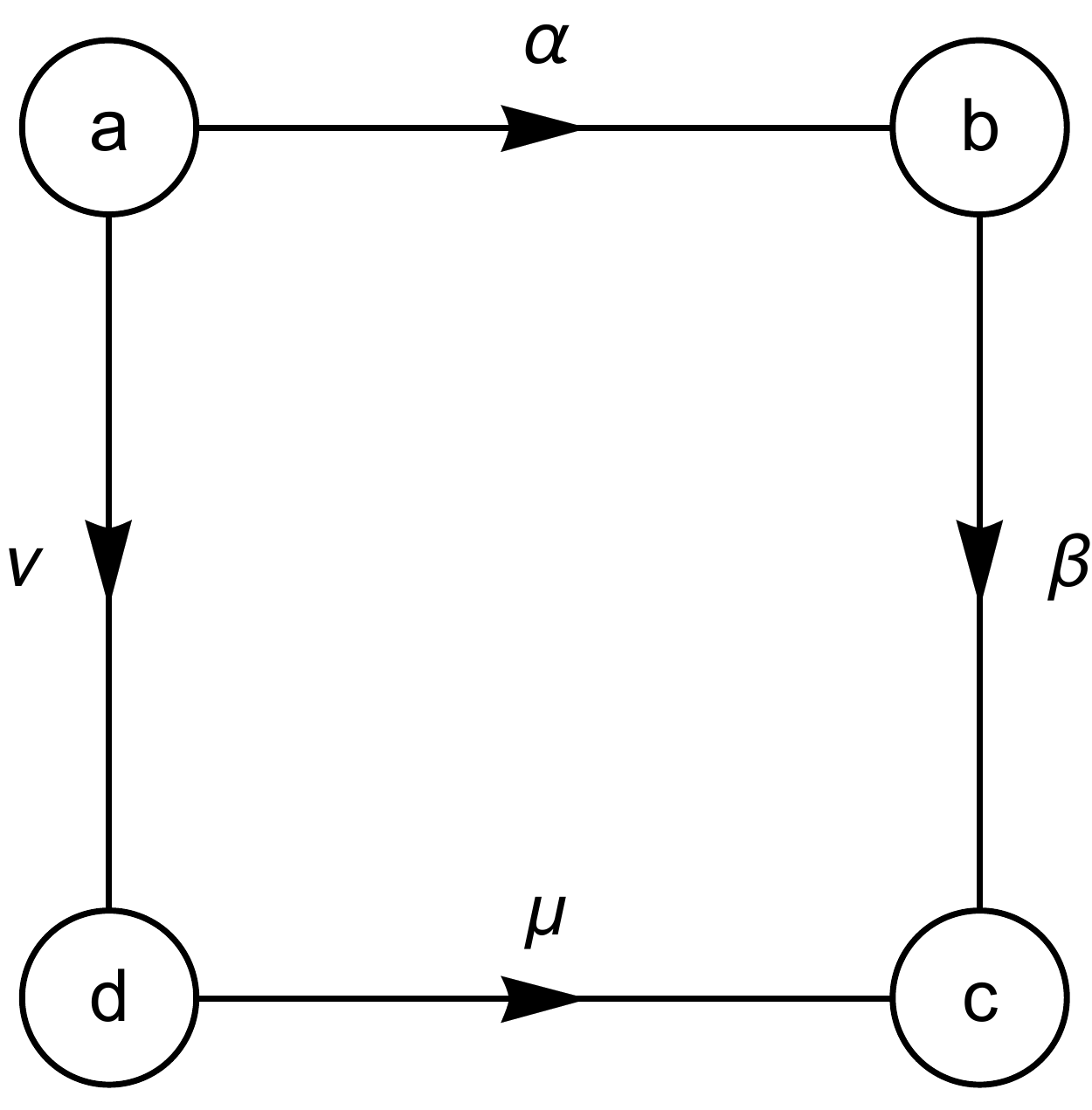}} ~ ~ ~
(b)~ \scalebox{0.3}[0.3]{\includegraphics{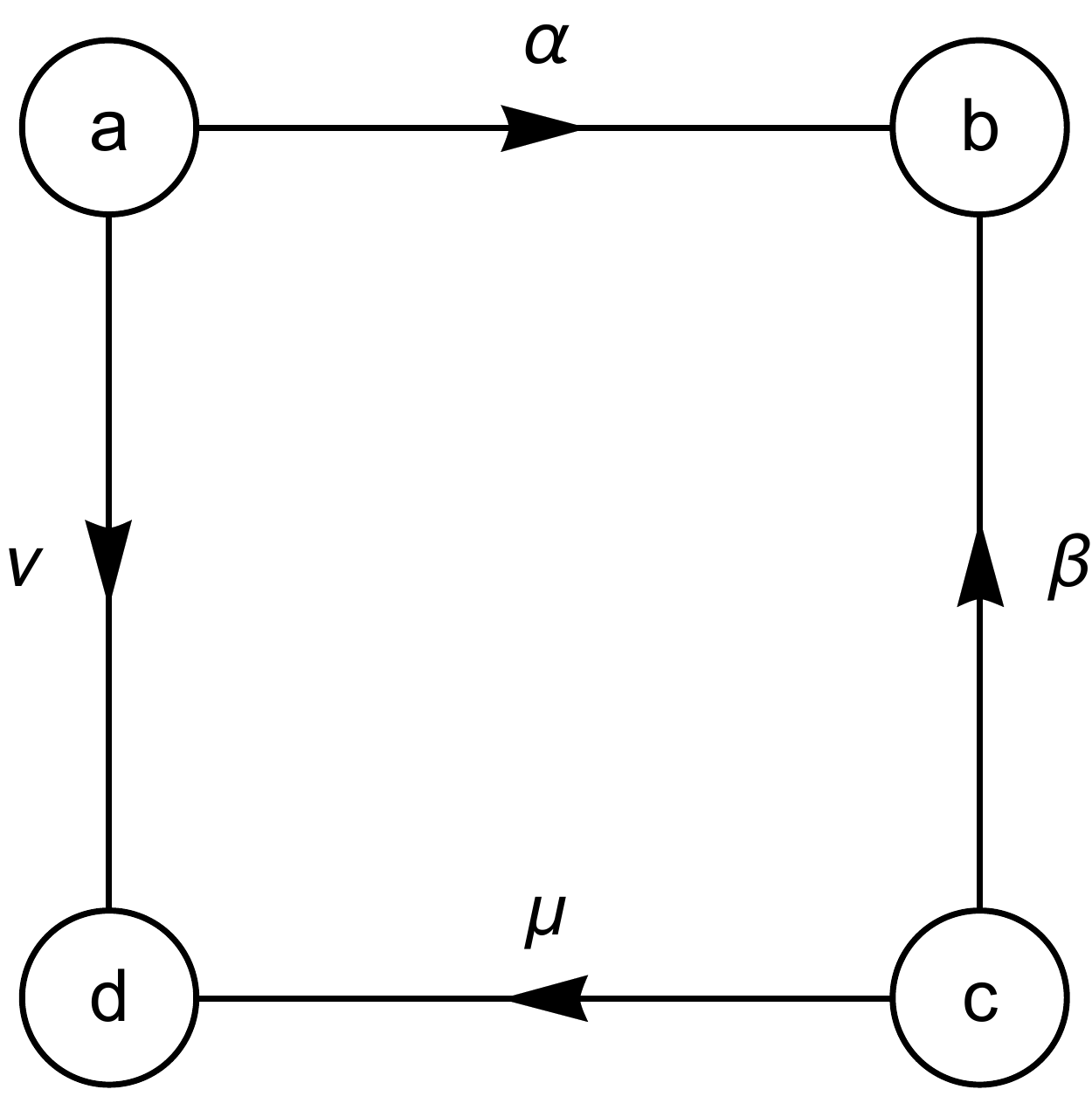}}
\caption{
Graphs of a 4-loop (a square), with two types of orientations: (a) non-bipartite; (b) bipartite.}
\label{fig:4-loop-graph}
\end{figure}

As the simplest example, let us explore a connectivity graph, generated by a length $4$ simple loop that consists of $4$ distinct edges, say, $\alpha = \{a, b\}$, $\beta = \{b, c\}$, $\mu = \{c, d\}$, and $\nu = \{d, a\}$, as shown in Fig.~\ref{fig:4-loop-graph}. We can call this graph a ``square''. We will first assume that this graph can be a part of a complex graph, then consider this graph as an entire graph. Our goal is to find restrictions on the 1-forms and LZ parameters that are imposed by Eqs.~(\ref{cycle-property-bar-A}) and (\ref{constr-3-LA-bar-A}). 

\subsection{Non-bipartite graph orientation}
We assume, initially, orientation to be arbitrary. By the  property (iii), $\bar{A}^{\alpha}$ and $\bar{A}^{\beta}$ are linearly independent and
\begin{eqnarray}
\label{4-loop-cycle-A} \bar{A}^{\nu} = x_{\alpha} \bar{A}^{\alpha} + x_{\beta} \bar{A}^{\beta}, \;\;\; \bar{A}^{\mu} = y_{\alpha} \bar{A}^{\alpha} + y_{\beta} \bar{A}^{\beta}.
\end{eqnarray}
We further make use of Eq.~(\ref{cycle-property-bar-A}) to
define the cyclic path
\begin{eqnarray}
\label{4-loop-cycle-n} n = s^{ab} \alpha + s^{bc} \beta + s^{cd} \mu + s^{da} \nu,
\end{eqnarray}
which is a cycle. Upon substitution of Eqs.~(\ref{4-loop-cycle-A}) and (\ref{4-loop-cycle-n}) into Eq.~(\ref{cycle-property-bar-A}), and looking at the coefficients in front of $\bar{A}^{\alpha} \otimes \bar{A}^{\alpha}$, $\bar{A}^{\beta} \otimes \bar{A}^{\beta}$, and $\bar{A}^{\alpha} \otimes \bar{A}^{\beta} + \bar{A}^{\beta} \otimes \bar{A}^{\alpha}$, we obtain a system of three quadratic equations
\begin{eqnarray}
\label{4-loop-cycle-eq-xy} s^{da}x_{\alpha}^{2} + s^{cd} y_{\alpha}^{2} + s^{ab} &=& 0, \nonumber \\ s^{da}x_{\beta}^{2} + s^{cd} y_{\beta}^{2} + s^{bc} &=& 0, \nonumber \\ s^{da} x_{\alpha}x_{\beta} + s^{cd} y_{\alpha}y_{\beta} &=& 0.
\end{eqnarray}

We will now consider all possible orientations that allow for nontrivial solutions of Eq.~(\ref{4-loop-cycle-eq-xy}). Let  $s^{da} = s^{cd}=1$. Then, for  Eq.~(\ref{4-loop-cycle-eq-xy}) to have nontrivial solutions we have $s^{ab} = s^{bc} = -1$, which means that $a$ and $c$ are a source and a sink, respectively, as we show in Fig.~\ref{fig:4-loop-graph}(a). Namely, they are, respectively, the origin and the destination of arrows that are connected to them. The other vertices, $b$ and $d$, are intermediate. We will call the whole orientation of edges in  Fig.~\ref{fig:4-loop-graph}(a) {\it non-bipartite}.
For this orientation, it is easy to see from Eq.~(\ref{4-loop-cycle-eq-xy}) that $(\bar{A}^{\nu}, \bar{A}^{\mu})$ are related to $(\bar{A}^{\alpha}, \bar{A}^{\beta})$ via an orthogonal ${\rm O}(2)$ transformation, i.e.,
\begin{eqnarray}
\label{A-O-transf} \bar{A}^{\nu} &=& \cos\varphi \bar{A}^{\alpha} + r \sin\varphi \bar{A}^{\beta}, \nonumber \\ \bar{A}^{\mu} &=& -\sin\varphi \bar{A}^{\alpha} + r \cos\varphi \bar{A}^{\beta},
\end{eqnarray}
with $r = \pm 1$, i.e., with the determinants of the corresponding $2 \times 2$ matrices equal $\pm 1$.
A  useful  
consequence of (\ref{A-O-transf}) is
\begin{eqnarray}
\label{A-O-determinants} \bar{A}^{\nu} \wedge \bar{A}^{\mu} = r \bar{A}^{\alpha} \wedge \bar{A}^{\beta},
\end{eqnarray}
which is true for a general ${\rm O}(2)$, not necessarily ${\rm SO}(2)$, transformation.

In what follows, it will be useful to view Eq.~(\ref{A-O-transf}) as a system of linear equations that relate different pairs of components of ${\bar A}$. Thus,
expressing $A^{\alpha}$ and $A^{\nu}$, via $A^{\mu}$ and $A^{\beta}$, we obtain
\begin{eqnarray}
\label{A-LO-transf-correspond} \bar{A}^{\alpha} &=& -\frac{1}{\sin\varphi} \bar{A}^{\mu} + r \frac{\cos\varphi}{\sin\varphi} \bar{A}^{\beta}, \nonumber \\ \bar{A}^{\nu} &=& -\frac{\cos\varphi}{\sin\varphi} \bar{A}^{\mu} + r \frac{1}{\sin\varphi} \bar{A}^{\beta},
\end{eqnarray}
and we further recast the result in the form of a pseudo-orthogonal transformation:
\begin{eqnarray}
\label{A-LO-transf-correspond-2} \bar{A}^{\alpha} &=& \tilde{p} \left(\cosh\vartheta \bar{A}^{\mu} + \tilde{r} \sinh\vartheta \bar{A}^{\beta}\right), \nonumber \\ \bar{A}^{\nu} &=& \tilde{p} \left(\sinh\vartheta \bar{A}^{\mu} + \tilde{r} \cosh\vartheta \bar{A}^{\beta}\right),
\end{eqnarray}
with the following relations:
\begin{eqnarray}
& \tilde{p} = - {\rm sgn} (\sin\varphi), \;\;\; \tilde{r} = -r, \;\;\; |\sin\varphi| \cdot \cosh\vartheta  = 1, \nn\\
& {\rm sgn}(\sinh\vartheta) = {\rm sgn}(\cos\varphi). \label{A-LO-transf-correspond-3}
\end{eqnarray}
An  analogue of Eq.~(\ref{A-O-determinants}) is now
\begin{eqnarray}
\label{A-LO-determinants-tilde} \bar{A}^{\alpha} \wedge \bar{A}^{\nu} = \tilde{r} \bar{A}^{\mu} \wedge \bar{A}^{\beta}.
\end{eqnarray}

By now we have assumed that this four-vertex graph can be a part of a complex graph. Let's now consider this four-vertex graph as an entire graph. In this case we can apply Eq.~(\ref{constr-3-LA-bar-A}), which leads to  two equations written for two pairs of opposite vertices, i.e., $\{a, c\}$ and $\{b, d\}$:
\begin{eqnarray}
 \sqrt{\gamma^{\alpha} \gamma^{\nu}}\bar{A}^{\alpha} \wedge \bar{A}^{\nu} & =& - \sqrt{\gamma^{\beta} \gamma^{\mu}}\bar{A}^{\beta} \wedge \bar{A}^{\mu}, \nn\\
 \sqrt{\gamma^{\nu} \gamma^{\mu}}\bar{A}^{\nu} \wedge \bar{A}^{\mu} &= & - \sqrt{\gamma^{\alpha} \gamma^{\beta}}\bar{A}^{\alpha} \wedge \bar{A}^{\beta}, \label{4-loop-graph-signs}
\end{eqnarray}
which can be reconciled with Eqs.~(\ref{A-O-determinants}) and (\ref{A-LO-determinants-tilde}) if we set
\begin{equation}
-r=\tilde{r}=1.
\label{r-cons}
\end{equation}
Eqs.~(\ref{A-O-determinants}), (\ref{A-LO-determinants-tilde}) and (\ref{4-loop-graph-signs}) also imply
\begin{eqnarray}
\label{4-loop-graph-gamma} \gamma^{\mu} = \gamma^{\alpha}, \;\;\; \gamma^{\nu} = \gamma^{\beta},
\end{eqnarray}
that is, the LZ parameters have to be the same for opposite links of the square graph.



\subsection{Bipartite graph orientation}

Another orientation that could produce a qualitatively different solution is shown in Fig.~\ref{fig:4-loop-graph}(b). This time,  both $b$ and $d$ are sinks of arrows, and both $a$ and $b$ are sources, and we refer to this graph orientation as {\it bipartite}.
  Eqs.~(\ref{4-loop-cycle-A}) define a pseudo-orthogonal transformation, whose general form has been already presented in Eq.~(\ref{A-LO-transf-correspond-2}), so that the forms $\bar A^{\sigma}$, associated with the edges of our $4$-loop are related by
\begin{eqnarray}
\label{A-LO-transf} \bar{A}^{\nu} &=& p \left(\cosh\vartheta \bar{A}^{\alpha} + r \sinh\vartheta \bar{A}^{\beta}\right), \nonumber \\ \bar{A}^{\mu} &=& p \left(\sinh\vartheta \bar{A}^{\alpha} + r \cosh\vartheta \bar{A}^{\beta}\right),
\end{eqnarray}
and we also have as a consequence Eq.~(\ref{A-O-determinants}) to hold. In the same way how Eq.~(\ref{A-LO-transf-correspond-2}) has been derived, we obtain
\begin{eqnarray}
\label{A-LO-transf-LO} \bar{A}^{\alpha} &=& \tilde{p} \left(\cosh\tilde{\vartheta} \bar{A}^{\beta} + \tilde{r} \sinh\tilde{\vartheta} \bar{A}^{\mu}\right), \nonumber \\ \bar{A}^{\nu} &=& \tilde{p} \left(\sinh\tilde{\vartheta} \bar{A}^{\beta} + \tilde{r} \cosh\tilde{\vartheta} \bar{A}^{\mu}\right),
\end{eqnarray}
with the following relations
\begin{eqnarray}
&\label{A-LO-transf-LO-3} \tilde{p} = - r \, {\rm sgn} (\sinh\vartheta), \;\;\; \tilde{r} = -r, \nn\\
&|\sinh\vartheta| \cdot |\sinh\tilde{\vartheta|} = 1, \;\;\; {\rm sgn}(\sinh\tilde{\vartheta}) = p.
\end{eqnarray}

Then we have for the bipartite orientation
\begin{eqnarray}
\label{connect-bipart} \bar{A}^{\nu} \wedge \bar{A}^{\mu} = \pm \bar{A}^{\alpha} \wedge \bar{A}^{\beta}, \quad \bar{A}^{\alpha} \wedge \bar{A}^{\nu} = \mp \bar{A}^{\beta} \wedge \bar{A}^{\mu},
\end{eqnarray}
where signs $\pm$ and $\mp$ are correlated with each other.
If this four-vertex graph is an entire graph, we will have \eqref{4-loop-graph-signs} as in the non-bipartite case. This means that Eqs.~(\ref{connect-bipart}) and (\ref{4-loop-graph-signs}) are contradictory to each other, so there is no nontrivial solution for the bipartite graph.

Therefore, 
for a 4-loop graph as an entire graph, the data (values of $\bar{A}_{\alpha}$ and $\gamma^{\alpha}$) on edges of this  graph satisfy integrability conditions only if its orientation is nonbipartite and conditions (\ref{A-O-transf}) with $r=-1$ (or, equivalently, (\ref{A-LO-transf-correspond-2}) with $\tilde{r}=1$) and (\ref{4-loop-graph-gamma}) are satisfied.

\subsection{Solvable 4-state models}
\label{sec:integr-cond-LF-examples-1}

Let us now construct an integrable model explicitly.
According to the directions of arrows in Fig.~\ref{4-loop-fig}, we have $\gamma^{12}, \gamma^{43}, \gamma^{14}, \gamma^{23}>0$. According to (\ref{A-LO-transf-correspond-2}), the relations between the $\bar A$ forms are:
\begin{align}\label{eq:pseudo}
\left(
  \begin{array}{c}
    \bar A^{34} \\
    \bar A^{23} \\
  \end{array}
\right)
=p  \left(\begin{array}{cc}
                 \cosh\vartheta  & \sinh\vartheta  \\
                 \sinh\vartheta  & \cosh\vartheta  \\
               \end{array}\right)
\left(
  \begin{array}{c}
    \bar A^{12} \\
    \bar A^{14} \\
  \end{array}
\right),
\end{align}
where $p=\pm 1$, and from \eqref{4-loop-graph-gamma} the relations for $\gamma^{ij}$ are:
\begin{align}\label{gamma-rel}
&\gamma^{12}=\gamma^{43},\quad \gamma^{14}=\gamma^{23}.
\end{align}

In what follows, to shorten notation, we will denote:
\be
 s\equiv \sinh\vartheta, \quad c \equiv \cosh\vartheta.
\label{cos-sin1}
\ee


Since the space of 1-forms is 2-dimensional, we can write
\be
 \bar{A}^{12} =  a_1dx^1+  a_2dx^2, \quad  \bar{A}^{14} =  {b}_1 dx^1+ {b}_2 dx^2,
\label{1-forms-any-2}
\ee
where $a_{1,2}$ and $ b_{1,2}$ are arbitrary real numbers. 

Using Eq.~(\ref{define-bar-A}) and identifying coefficients of $A^{ij}$ near $dx^1$ with couplings in $H_1$ we find
\be
&g_{12}=\sqrt{\gamma^{12}} a_1, \quad g_{14}=\sqrt{\gamma^{14}} b_1, \quad
g_{23} = p\sqrt{\gamma^{14}}(sa_1+cb_1), \nn\\
& g_{34}=p\sqrt{\gamma^{12}}(ca_1+sb_1).
\label{coeff1-sol1}
\ee
Equation~(\ref{coeff1-sol1}) defines four couplings in terms of five free parameters of the model: $a_1$, $b_1$, $\vartheta$, $\gamma^{14}$ and $\gamma^{12}$. So, we have freedom to set the couplings to arbitrary different values (with one exception to which we will return).

However, slopes of the levels are generally not independent. Recalling Eqs.~(\ref{constr-2-LA-3}) and  (\ref{constr-3-LA}), and identifying $\Lambda_{11}^{s}$ with $\beta_s$ in (\ref{h-four-ex}), we find
\be
 & \beta_1-\beta_2=a_1^2, \quad \beta_1-\beta_4=b_1^2, \quad \beta_2-\beta_3=(sa_1+cb_1)^2, \nn\\
& \beta_4-\beta_3=(ca_1+sb_1)^2.
\label{betas-find}
\ee
Equations in (\ref{betas-find}) are dependent on each other because they give identity if we sum all of them with proper signs. This merely reflects the freedom to do a gauge transformation
\be
H_1 \rar H_1+(\beta t+e) \hat{1}
\label{gauge1}
\ee
that keeps the Hamiltonian integrable. Apart form this, there  are no new free parameters that resolve Eq.~(\ref{betas-find}). Finally, using Eqs.~(\ref{constr-2-LA-3})-(\ref{constr-3-LA}) and identifying $\Lambda_{12}^{s}$ with $e_s$ in (\ref{h-four-ex}), we find
\be
e_1-e_2=a_1 a_2, \quad e_1-e_4=b_1b_2, \nn\\
\quad e_2-e_3=(sa_1+cb_1)(sa_2+cb_2), \nn\\
\quad e_4-e_3=(ca_1+sb_1)(ca_2+sb_2).
\label{es-find}
\ee
Again, this set of equations determines $e_s$, $s=1,2,3,4$, up to a gauge freedom constant in (\ref{gauge1}). Note, however, that this is the only place where the new free parameters, $a_2$ and $b_2$ appear. Hence, unlike the slopes $\beta_s$, the parameters  $e_s$  are not completely determined by the values of model's couplings.

Summarizing, we found simple equations~(\ref{coeff1-sol1}), (\ref{betas-find}), and (\ref{es-find}), that determine all parameters of the Hamiltonian (\ref{h-four-ex}) up to the gauge freedom (\ref{gauge1}).
Thus, the resulting model depends on seven {\it independent} real parameters: $a_{1,2}$, $b_{1,2}$, $\vartheta$, $\gamma^{12}$, and $\gamma^{14}$. We can also add to this list the sign index $p$ in  (\ref{coeff1-sol1}).

At this stage, the Hamiltonian $H_1$ does not look particularly ``physical". However, this Hamiltonian does have a simple physical interpretation if  two couplings are set the same, e.g., let
\be
g_{12}=g_{34}=g_1,
\label{ddeq}
\ee
where $g_1$ is an arbitrary constant.
There are two choices of free parameters at which this occurs.
The first choice is the case with
\be
\vartheta=0.
\label{varth0}
\ee
We find then that the Hamiltonian can be parametrized so that
\begin{align}\label{h-four-ex2}
 H_1(t)=\left( \begin{array}{cccc}
\beta_1 t+e_1 & g_1 & 0 & g_2 \\
g_1 & \beta_2 t+e_2 &  g_2&  0 \\
0 & g_2 &  \beta_3 t+e_3  & g_1 \\
g_2  & 0 & g_1  & \beta_4 t+e_4  \\
\end{array}\right),
\end{align}
where the only constraints on the parameters are
$$
\beta_1-\beta_2= \beta_3-\beta_4, \quad \beta_1-\beta_4=\beta_2-\beta_3,
$$
and
$$
e_1-e_2=e_4-e_3, \quad e_1-e_4=e_2-e_3,
$$
i.e., the couplings  $g_1$ and $g_2$ are independent of the diagonal elements.  This particular choice is trivial. It coincides with the Hamiltonian
\be
H_1=\hat{1}_2\otimes {\cal H}_1^{LZ} + {\cal H}_2^{LZ} \otimes \hat{1}_2,
\label{indep-h}
\ee
that describes two noninteracting spins experiencing independent two-state LZ transitions that are described by independent 2$\times$2 Hamiltonians ${\cal H}_{1,2}^{LZ}$.
This trivial case was discussed previously in \cite{multiparticle}.

A nontrivial case is found if we set $\vartheta \ne 0$, i.e. $c\ne 1$. Then substituting (\ref{ddeq}) into (\ref{coeff1-sol1}) we find
\be
g_{14}=-g_{23}=g_1,
\label{ggeq}
\ee
i.e., this is a special case at which the couplings $g_{14}$ and $g_{23}$ cannot be made arbitrary. However, specifically at this case, Eq.~(\ref{betas-find}) does not have unique resolution. Up to
a shift of time $t\rar t+t_0$, the Hamiltonian (\ref{h-four-ex}) can then be parametrized
as follows:
\begin{align}\label{h-four-ex3}
 H_1(t)=\left( \begin{array}{cccc}
\beta_1 t+e_1 & g_1 & 0 & g_2 \\
g_1 & \beta_2 t -e_1 &  -g_2&  0 \\
0 & -g_2 &  -\beta_1 t+e_1  & g_1 \\
g_2  & 0 & g_1  & - \beta_2 t-e_1  \\
\end{array}\right),
\end{align}
where all parameters are independent. Comparing this Hamiltonian with the Hamiltonian (\ref{ham-string1}) for $N=2$, we find that up to renaming of variables they are the same.
Thus, as expected, the $N=2$ case of the $\gamma$-magnet model is a special case of the square graph family.

One can easily construct a commuting Hamiltonian for (\ref{h-four-ex3}) by identifying couplings with coefficients of ${A}^s$ at $dx^2$ and so on. Since we already proved that the square-family is 2-dimensional, we also proved that
the Hamiltonian (\ref{ham-string1}) for $N=2$ does not have other nontrivial operators but (\ref{ham-string2}). 


\section{Cube}
\label{sec:integr-cond-LF-examples-cube}

Let us now extend the analysis of a simple square graph to  an $8$-state MTLZ model whose graph is a cube, as shown in Fig.~\ref{fig:graph-cube}. A specific case of this model was considered in \cite{gammamagnet}. Here we will consider its most general form. 
\begin{figure}[!htb]
\scalebox{0.3}[0.3]{\includegraphics{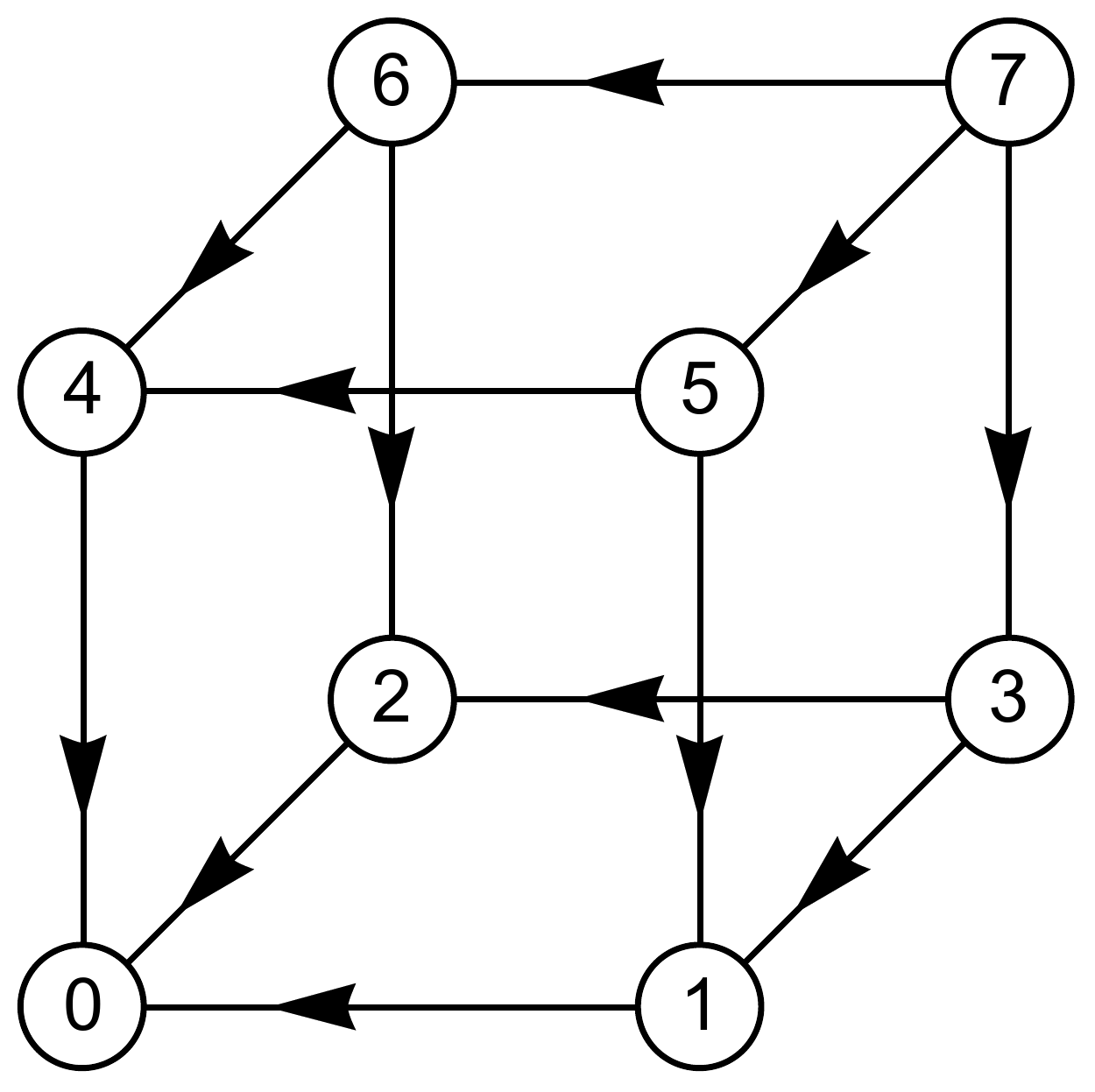}}
\caption{
Directed graph of the cube model, where every 4-loop has a non-bipartite orientation.}
\label{fig:graph-cube}
\end{figure}

\subsection{MTLZ family on cube}
The cube graph is shown in Fig.~\ref{fig:graph-cube}. It has the property that for any two vertices that can be connected by a length-2 path (namely, any two vertices that sit on diagonal position of one face of the cube), there are only two such paths in the entire graph. Thus, the graph is decomposable into 4-loops. According to the analysis in the previous section for a square model, all these 4-loops should have non-bipartite orientations. Therefore, up to a permutation of vertices, we get only one type of orientations, as shown in Fig.~\ref{fig:graph-cube}. Vertex 0 is a sink, vertex 7 is a source, and all other vertices are intermediate.

For $\gamma^{ab}$, considering loop 0132, an argument similar to that for the square model gives:
\begin{align}\label{}
&|\gamma^{01}|=|\gamma^{23}|,\quad |\gamma^{02}|=|\gamma^{13}|.
\end{align}
Writing out similar relations for all other 4-loops, we see that the twelve LZ parameters $\gamma^{ab}$'s are related 
so  that any four edges parallel to each other have the same $|\gamma^{ab}|$. There are only three values of $|\gamma^{ab}|$ that become the independent parameters. Including the signs determined by $s^{ab}$'s, which are illustrated by the arrows in Fig.~\ref{fig:graph-cube}, we get
\begin{align}\label{}
&\gamma^{01}=\gamma^{23}=\gamma^{45}=\gamma^{67},\\
&\gamma^{02}=\gamma^{13}=\gamma^{46}=\gamma^{57},\\
&\gamma^{04}=\gamma^{15}=\gamma^{26}=\gamma^{37},
\end{align}
and that all these twelve $\gamma^{ab}$s are positive. 

For $\bar A^{ab}$, we use the previous result that for any 4-loop its four $\bar A^{ab}$ forms are connected via orthogonal or pseudo-orthogonal transformations. Let's take the three forms on edges connected to vertex 0, namely, the forms $\bar A^{01}$, $\bar A^{02}$ and $\bar A^{04}$, to be known. For simplicity, later we will write the loop indices as a single number, by making the substitutions
\be
&0132\rar 1, \quad 0154\rar 2, \quad 0264\rar 3, \nn\\
&1375\rar 4, \quad 2376\rar 5, \quad 4576\rar 6. \nn
\ee
For the three loops that include vertex 0, namely the loops 0132, 0154, and 0264, the pseudo-orthogonal transformations give:
\begin{align}
&\left(\begin{array}{c}
\bar A^{23} \\
\bar A^{13} \end{array} \right) = U(\vartheta_{1})\left( \begin{array}{c}
\bar A^{01}\\
\bar A^{02} \end{array} \right), \quad
\left( \begin{array}{c}
\bar A^{45}\\
\bar A^{15} \end{array} \right)=  U(\vartheta_{2}) \left( \begin{array}{c}
\bar A^{01}\\
\bar A^{04} \end{array} \right), \nn\\
&\left( \begin{array}{c}
\bar A^{46} \\
\bar A^{26} \end{array} \right)=  U(\vartheta_{3}) \left( \begin{array}{c}
\bar A^{02}\\
\bar A^{04} \end{array} \right),\label{loops1}
\end{align}
with
\begin{align}\label{}
U(\vartheta_{i})=p_i\left(
               \begin{array}{cc}
                 \cosh \vartheta_i & \sinh \vartheta_i \\
                 \sinh \vartheta_i & \cosh \vartheta_i \\
               \end{array}
            \right)\equiv
		\left(
               \begin{array}{cc}
                 c_i & s_i \\
                 s_i & c_i \\
               \end{array}
            \right),
	\end{align}
where $p_i =\pm1$ are sign factors, and
\begin{align}\label{}
&c_i=p_i\cosh\vartheta_i,\quad s_i=p_i\sinh\vartheta_i, \quad c_i^2-s_i^2=1.
\end{align}
$U(\vartheta_{i})$ is a pseudo-orthogonal matrix, with $\vartheta_i$'s being rapidities which can take values from $-\infty$ to $\infty$. The other three loops 
which include vertex 7 then give:
\begin{align}\label{loops2}
&\left(\begin{array}{c}
\bar A^{57} \\
\bar A^{37} \end{array} \right) = U(\vartheta_{4})\left( \begin{array}{c}
\bar A^{13}\\
\bar A^{15} \end{array} \right), \quad
\left( \begin{array}{c}
\bar A^{67}\\
\bar A^{37} \end{array} \right)=  U(\vartheta_{5}) \left( \begin{array}{c}
\bar A^{23}\\
\bar A^{26} \end{array} \right), \nn\\
&\left( \begin{array}{c}
\bar A^{67} \\
\bar A^{57} \end{array} \right)=  U(\vartheta_{6}) \left( \begin{array}{c}
\bar A^{45}\\
\bar A^{46} \end{array} \right).
\end{align}


Equations~(\ref{loops2}) overdetermine the forms $\bar A^{37}$, $\bar A^{57}$ and $\bar A^{67}$ because there are two equations for each of them. For example, substituting (\ref{loops1}) into (\ref{loops2}) we find two expressions for $\bar A^{37}$ in terms
of the forms that we consider linearly independent:
\begin{align}\label{}
&\bar A_{37}= c_{4} ( c_{2}\bar A^{04} +
       s_{2}\bar A^{01} ) +
	  s_{4}(c_{1}\bar A^{02}  +s_{1}
       \bar A^{01} ),\label{eq:constraint-A37a}\\
&\bar A_{37}=
 c_{5} (c_{3}\bar A^{04}  +
       s_{3}\bar A^{02}) +
   s_{5}  (c_{1}\bar A^{01}  +
       s_{1} \bar A^{02}) .\label{eq:constraint-A37b}
\end{align}
If we assume that the three forms $\bar A^{01}$, $\bar A^{02}$ and  $\bar A^{04}$ are linearly independent, then the coefficients near these forms in Eqs.~\eqref{eq:constraint-A37a} and \eqref{eq:constraint-A37b} should be the same, which gives three conditions on $\vartheta$s:
\begin{align}\label{}
&c_{2}   c_{4}=c_{3}  c_{5},\label{eq:consistency-condition-1}\\
& c_{1}s_{5} =s_{2}  c_{4} + s_{1}s_{4} ,\label{eq:consistency-condition-2}\\
&c_{1}s_{4} = s_{3}c_{5}+s_{1} s_{5} .\label{eq:consistency-condition-3}
\end{align}
This is a system of three equations with five variables ($s_i$ and $c_i$ are viewed as the same variable), but one equation turns out to follow from the other two. Therefore, there are three rapidities that we can consider as independent parameters of the model and
derive other rapidities from them.

Let us simplify the information that is contained in Eqs.~\eqref{eq:consistency-condition-1}-\eqref{eq:consistency-condition-3}. From Eq.~\eqref{eq:consistency-condition-1}, we have $c_5=c_2 c_4/c_3$. From Eq.~\eqref{eq:consistency-condition-2}, we have $s_5=(s_{2}  c_{4} + s_{1}s_{4})/c_1$. Plugging these two expressions  into Eq.~\eqref{eq:consistency-condition-3}, we get:
\begin{align}\label{}
&c_{1}s_{4} = \frac{c_2 c_4}{c_3} s_{3}+\frac{s_{2}  c_{4} + s_{1}s_{4}}{c_1} s_{1},
\end{align}
which is equivalent to
\begin{align}\label{eq:s2c2}
&\frac{s_{4}}{c_{4}} = \frac{c_1c_2 s_3+s_1 s_2 c_3}{c_3},
\end{align}
where we  used $c_1^2-s_1^2=1$.

Let's now introduce the hyperbolic tangents:
\begin{align}\label{}
\tau_i=\frac{s_i}{c_i}=\tanh\vartheta_i.
\end{align}
In terms of $\tau_i$, the functions $s_i$ and $c_i$ are expressed as:
\begin{align}\label{sc-in-tau}
&s_i=\frac{p_i\tau_i}{\sqrt{1-\tau_i^2}},\quad c_i=\frac{p_i}{\sqrt{1-\tau_i^2}}.
\end{align}
Plugging these into Eq.~\eqref{eq:s2c2}, we get an expression of $\tau_4$ in terms of $\tau_1$, $\tau_2$ and $\tau_3$:
\begin{align}\label{eq:tau2}
&\tau_4 = \frac{p_1p_2(\tau_3  +\tau_1\tau_2)}{\sqrt{(1-\tau_1^2)(1-\tau_2^2)}}.
\end{align}

Now we note that our graph in Fig.~\ref{fig:graph-cube} possesses a $3$-fold rotation symmetry about the line connecting vertices 1 and 7. Therefore, the expressions for $\bar A^{67}$ and $\bar A^{78}$ can be directly obtained from those for $\bar A^{37}$ (Eqs.~\eqref{eq:constraint-A37a} and \eqref{eq:constraint-A37b}) by exchanges of indices according to this symmetry.Thus, we find the expressions for $\tau_5$ and $\tau_6$ in terms of $\tau_1$, $\tau_2$ and $\tau_3$:
\begin{align}
&\tau_5 = \frac{p_1p_3(\tau_2  +\tau_1\tau_3)}{\sqrt{(1-\tau_1^2)(1-\tau_3^2)}},\label{eq:tau4}\\
&\tau_6 = \frac{p_2p_3(\tau_1  +\tau_2\tau_3)}{\sqrt{(1-\tau_2^2)(1-\tau_3^2)}}\label{eq:tau5}.
\end{align}
The sign factors $p_i$ are also not all independent, and they satisfy:
\begin{align}
p_2 p_4=p_3p_5,\quad p_1 p_4=p_3p_6.
\end{align}

Note that in the analysis above we assumed linear independence of $\bar A^{01}$, $\bar A^{02}$ and  $\bar A^{04}$, and the space of the $\bar A$ forms is 3-dimensional. We did also consider the case when not all of $\bar A^{01}$, $\bar A^{02}$ and  $\bar A^{04}$ are linearly independent. Then the dimension of the space of the $\bar A$ forms has to be 2 due to the good family assumption. In this case, all the rapidities can be taken as independent parameters, and we tried to solve for all the $\bar A$ forms, but we found that the equations always lead to a contradiction. This indicates that there are no intrinsic 2-dimensional families on the cube graph. Namely, any 2-dimensional family on  the cube can be obtained trivially from a 3-dimensional family that we just described by restricting to a 2-dimensional subspace.

Summarizing, we found that the cube connectivity graph describes a 3-dimensional MTLZ family, which is parametrized by nine parameters of three independent 1-forms: $\bar{A}^{01}$, $\bar{A}^{02}$, and $\bar{A}^{04}$ plus three independent LZ parameters: $\gamma^{01}$, $\gamma^{02}$ and $\gamma^{04}$ plus three independent rapidity parameters, or rather their hyperbolic tangents: $\tau_1$, $\tau_2$ and $\tau_3$, whose values should keep other
such variables, $\tau_4$, $\tau_5$ and $\tau_6$ within the range $(-1,1)$.  There is one trivial choice of the rapidities:
$\tau_i=0$ for all $i$.
We verified that this case corresponds to a trivial  model that is composed of three independent 2$\times$2 LZ Hamiltonians:
 \be
 H_{{\bm \tau}=0} &= & {\cal H}_1^{LZ} \otimes \hat{1}_2 \otimes \hat{1}_2 + \hat{1}_2 \otimes {\cal H}_2^{LZ} \otimes \hat{1}_2 \nn\\
 &  +  & \hat{1}_2\otimes \hat{1}_2 \otimes {\cal H}_3^{LZ}.
 \label{composed3}
 \ee
The transition probability matrix \cite{large-class} is then a direct product of three 2$\times$2 LZ probability matrices:
\begin{align}\label{eq:P-0}
 &P_{{\bm \tau}=0} = {P}_1^{LZ} \otimes {P}_2^{LZ} \otimes{P}_3^{LZ},\nn\\
 &{P}_i^{LZ}=\left(\begin{array}{cc}
 p_i  & q_i \\
 q_i & p_i 
 \end{array}\right),\quad (i=1,2,3),
 \end{align}
 where $p_1=e^{-2\pi\gamma^{01}}$, $p_2=e^{-2\pi\gamma^{02}}$, $p_3=e^{-2\pi\gamma^{04}}$, $q_1=1-p_1$, $q_2=1-p_2$, and $q_3=1-p_3$.
(Note that here $p_i$ are probabilities instead of sign factors, although we use the same notation for both.)

Are there nontrivial cases in addition to \eqref{composed3}? The answer is yes -- at least one such case, the $\gamma$-magnet, has been found \cite{gammamagnet}. The connectivity graph for the $\gamma$-magnet \cite{gammamagnet} with $N$ spins is the $N$-dimensional hypercube, and it is a cube at $N=3$. 
Given the large set of parameters described above, it is natural to ask whether there are more solutions on the cube graph. In the next subsection, we  show that the family of solutions on cube is actually very rich.



\subsection{Classification of solutions on cube}

Let us now outline the strategy to classify different Hamiltonians that correspond to different transition probability matrices within the same cube family.
In what follows, we assume that the reader is familiar with section~8 from Ref.~\cite{large-class}.

According to \cite{commute}, for a MTLZ model \eqref{system1}, if we choose a linear time path via the substitution
\begin{align}\label{time-path}
\mathcal{P}_t:\;x^i (t) = v^i t + \varepsilon^i,\;\;\; {\rm for}\;\; i=1,\ldots, M
\end{align}
with arbitrary parameters $v^i$ and $\varepsilon^i$, then \eqref{system1} reduce to a multistate LZ model \eqref{multistate LZ} with the Hamiltonian
\begin{align}
H(t) = v^i H_i(x^1(t),\ldots,x^M(t)).
\end{align}
This property provides a way to generate multistate LZ Hamiltonians. 
According to \cite{commute,large-class}, a scattering problem for the MTLZ model can be solved by a WKB-like approach. For evolution along the path \eqref{time-path} from $t=-\infty$ to $t=\infty$, the path can be deformed to a path $\mathcal{P}_\infty$ along which $|\mathbf{x}|$ is always large. The path $\mathcal{P}_\infty$ goes through a series of adiabatic regions, within which the adiabatic energy levels are well separated. These adiabatic regions are separated by hyperplanes which correspond to pairwise degeneracies of the diabatic energy levels of the Hamiltonians $H_i$. The positions of these hyperplanes are determined by the conditions $\bar A^{ab}_{j}x^j = 0$. 
We can label a hyperplane by the indices $ab$. 

When $\mathcal{P}_\infty$ goes across the hyperplane $ab$, the scattering matrix experiences a jump described by a ``connecting matrix''. It is a unit matrix except for the $2\times2$ block for the levels $a$ and $b$, which coincides with a scattering matrix for a $2\times2$ LZ model where $\gamma^{ab}$ enters as a parameter (see Eq.~(87) in \cite{large-class}). The direction when $\mathcal{P}_\infty$ crosses the hyperplane $ab$ also influences the connecting matrix -- if we denote the connecting matrix when $\mathcal{P}_\infty$ goes from a $\bar A^{ab}_{j}x^j>0$ region to a $\bar A^{ab}_{j}x^j<0$ region as $S^{ab}$, then the connecting matrix will become $(S^{ab})^\dag$ when $\mathcal{P}_\infty$ takes the opposite direction. The scattering matrix of the whole evolution is then a product of a series of adiabatic evolution matrices and LZ matrices ordered along the path $\mathcal{P}_\infty$.  It has been shown \cite{commute,large-class} that the adiabatic evolution matrices produce phase factors that always cancel out in the expressions of transition probabilities for the whole evolution, and the connecting matrices completely determine the transition probabilities. Here we will apply this approach to the cube graph.


Generally, each of the $3$ independent 1-forms $\bar{A}^{01}$, $\bar{A}^{02}$, and $\bar{A}^{04}$ will have three arbitrary components in $dx^1$, $dx^2$ and $dx^3$. We will define new coordinates $dx^1$, $dx^2$ and $dx^3$ such that
\begin{align}
\bar{A}^{01}=dx^1,\quad \bar{A}^{02}=dx^2,\quad \bar{A}^{04}=dx^3. 
\end{align}
This corresponds to performing a linear transformation on the coordinate system. Now the three 1-forms carry no free parameters but parameters $v^i$ and $\varepsilon^i$ from (\ref{time-path}) are used instead.
After this transformation, all the $\bar A^{ab}$ forms are completely determined by the rapidities $\tau_1$, $\tau_2$, $\tau_3$ (or, more precisely, the rapidities and the sign factors $p_i$), so the positions of hyperplanes depend only on the rapidities and not on other parameters in the list. The parameters $v^i$ ($i=1,2,3$), on the other hand, determine which adiabatic regions the evolution starts from and end with. The parameters $\varepsilon^i$ ($i=1,2,3$) give shifts to the energy constants on the diagonal entries of $H(t)$, and they do not affect transition probabilities. Finally, the LZ parameters $\gamma^{01}$, $\gamma^{02}$ and $\gamma^{04}$ enter only as parameters of the connecting matrices and determine values of transition probabilities, but they do not influence the structure of the transition probability matrix.

Let us now sketch how one can perform the classification of different behavior within the family of solvable models. We first make a choice of $\tau_1$, $\tau_2$, and $\tau_3$, and calculate all the $\bar A^{ab}$ forms using Eqs.~\eqref{loops1}, \eqref{loops2}, and (\ref{sc-in-tau})-(\ref{eq:tau5}). We then find the position of any  hyperplane $ab$ by solving $\bar A^{ab}_{j}x^j = 0$. For a 3-dimensional MTLZ family like the cube model, the  hyperplanes are 2D planes passing the origin of the 3D space spanned by $x^1$, $x^2$, and $x^3$.
If we draw a sphere $S^2$ in this 3D space, these planes will intersect the sphere along great circles. We will label a great circle also by $ab$. Each great circle $ab$ separates the sphere into two hemispheres, one with $\bar A^{ab}_{j}x^j >0$, and the other with $\bar A^{ab}_{j}x^j <0$.

 Since there are  twelve different nonzero $\bar A^{ab}$ forms, there are twelve such great circles. Altogether, they decompose the sphere $S^2$ into a number of cells, and each cell corresponds to an adiabatic region. Let's now choose the radius of $S^2$ to be large. Recall that we are considering evolution along the path $\mathcal{P}_t$ (Eq.~\eqref{time-path}) which is deformed to $\mathcal{P}_\infty$. The evolution path $\mathcal{P}_t$ intersect the sphere $S^2$ (with a large radius) at two points which lie in two cells. We will call them the  initial and final cells for a given evolution path. (Note that, on the sphere $S^2$, these two cells are always at positions opposite to each other.) Once we make a choice of $v^i$ ($i=1,2,3$), the initial and final cells are fixed. We then deform $\mathcal{P}_t$ to $\mathcal{P}_\infty$ while keeping its two intersecting points with the sphere $S^2$ fixed. $\mathcal{P}_\infty$ can be chosen to run on $S^2$, where it becomes a path threading a number of cells. Adiabatic evolution takes place within a cell, but not when it goes from one cell to another. 
 
 Consider now a segment of $\mathcal{P}_\infty$ that connects two neighboring cells separated by the great circle $ab$. When going along this segment, $\bar A^{ab}_{j}x^j$ changes sign, 
and evolution along this segment contributes to the scattering matrix a connecting matrix $S^{ab}$ or $(S^{ab})^\dag$ with the parameter $\gamma^{ab}$, as described in Ref.~\cite{large-class}. We can then choose a path that connects the initial and final cells, and write all the connecting matrices between the neighboring cells along this path, and then obtain the transition probability matrix for the whole evolution. The way to choose this path is not unique but the final scattering matrix does not depend on this choice \cite{commute}. We also note that if $v_i$ is changed but the initial cell remains the same, then the final scattering matrix also remains the same, since evolution within a cell is adiabatic. Thus, the choice of parameters $v_i$ is reduced to a choice of the initial cell. 

\begin{figure}[!htb]
\smallskip
 \scalebox{0.75}[0.75]{\includegraphics{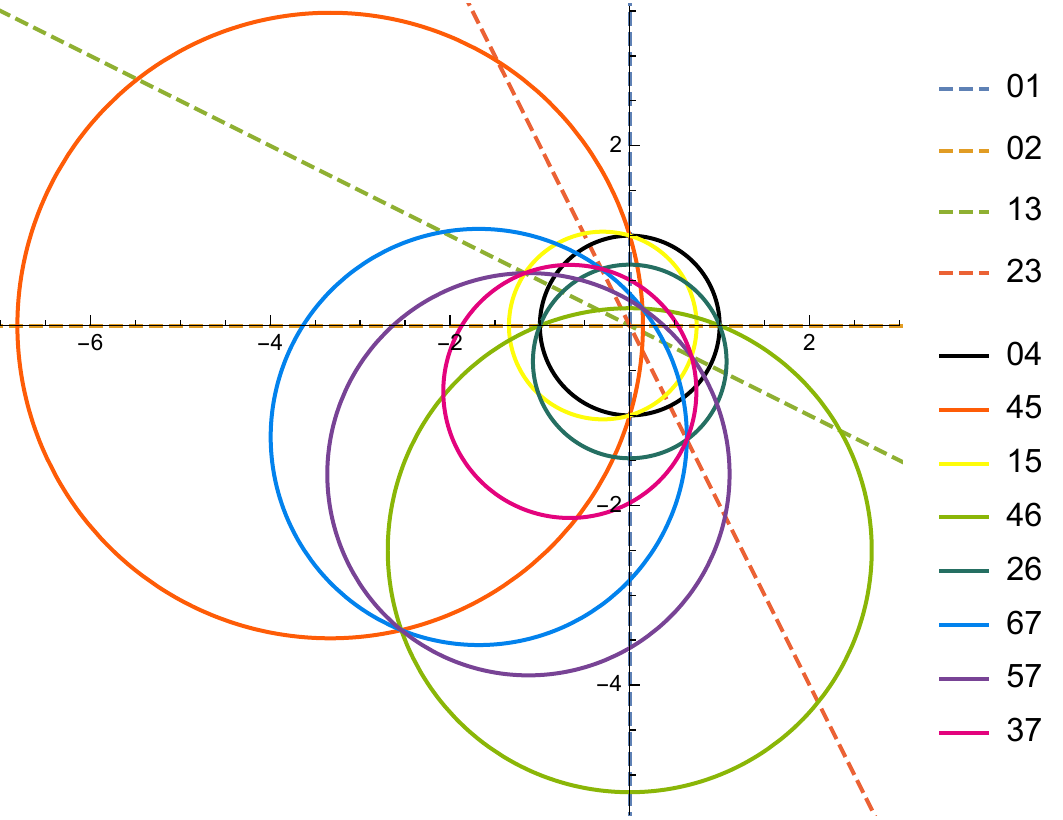}}
\caption{An example of the cell decomposition plot, which is stereographic-projected on a plane. The parameters are: $\tau_1= 0.5$, $\tau_2 = 0.3$, $\tau_3=0.4$, and all sign factors $p_i=1$. The label for a circle or a straight line in the legend is the same as the label $ab$ for a great circle being the solution to $\bar A^{ab}_j x^j=0$, (e.g. 01 corresponds to the great circle with $\bar A^{01}_j x^j=0$).}
\label{fig:cell-decomposition}
\end{figure}

Figure~\ref{fig:cell-decomposition} is an example of a cell decomposition plot for some choice of $\tau_1$, $\tau_2$ and $\tau_3$ on the cube geometry. To show the decomposition in a planar figure, we perform a stereographic projection which transforms a sphere $S^2$ to a plane. A great circle on the sphere then transforms either to a circle or a straight line on the plane. There are totally ninety eight cells, and each of them can be chosen as the initial cell of an evolution. Once we computed transition probabilities of all these evolutions, we find all possible solutions at a given choice of $\tau_1$, $\tau_2$ and $\tau_3$. 

For computing transition probabilities, it is convenient to draw a graph dual to the cell decomposition plot. In such a dual graph, each cell is represented by a vertex, and each pair of two neighboring cells are connected by an edge, which is dual to the segment of the great circle that separates the two cells. If that great circle is labelled by $ab$, we will associate to the edge the connecting matrix $S^{ab}$ (which is a function of $\gamma^{ab}$). We also define an orientation on each edge  -- on an edge crossing the big circle $ab$, we put an arrow which points from the $\bar A^{ab}_{j}x^j>0$ side to the $\bar A^{ab}_{j}x^j<0$ side. On the dual graph, a path of an evolution becomes a path of connected edges. Each edge contributes to the overall scattering matrix a factor $S^{ab}$ (or $(S^{ab})^\dag$) if the path goes in (or opposite to) the direction of the arrow. We can then directly read out the series of connecting matrices for that evolution. Besides, the dual graph reveals symmetric structures of the cells, which allow us to calculate the transition probabilities for only a portion of the choices of initial cells and obtain the transition probabilities for the remaining choices by symmetry.

We worked out cell decompositions and the corresponding dual graphs for several different choices of $\tau_1$, $\tau_2$ and $\tau_3$, with all sign factors $p_i$ positive. We considered three cases whose structures of cell decompositions were different: 1) When all $\tau_i$ ($i=1,\ldots, 6$) are positive (the cell decomposition in Fig.~\ref{fig:cell-decomposition} belongs to this case). 2) When $\tau_1<0$, $\tau_2,\tau_3>0$ and $\tau_4,\tau_5,\tau_6>0$ (namely, when one $\tau_i$ is negative). 3) When $\tau_1<0$, $\tau_2,\tau_3>0$, $\tau_4<0$ and $\tau_5,\tau_6>0$ (namely, when two $\tau_i$'s are negative).

A simultaneous change of signs of two of $\tau_1$, $\tau_2$ and
$\tau_3$ leads to sign changes of two of $\tau_4$, $\tau_5$ and $\tau_6$ and
leaves their amplitudes unchanged, as can be seen from Eqs.~\eqref{eq:tau2}-\eqref{eq:tau5}. This fact results in the cell decomposition plot to be just a reflection of the plot before the simultaneous sign change. For example, for the choice $\tau_1 =- 0.5$, $\tau_2 =- 0.3$ and
$\tau_3 = 0.4$, which is the choice in Fig.~\ref{fig:cell-decomposition} with the signs of $\tau_1$ and $\tau_2$
flipped, the cell decomposition plot becomes a reflection of Fig.~\ref{fig:cell-decomposition} about the vertical axis. We checked that all transition probability matrices remain unchanged as compared to those before the flips of signs of $\tau_1$ and $\tau_2$. Since all choices of $\tau_1$, $\tau_2$ and $\tau_3$ can be connected to either one of $\tau_1$, $\tau_2$ and $\tau_3$ being negative or all of them being
positive by such a simultaneous flip, the three cases described above should eliminate all possibilities of values of $\tau_i$. We also checked a case when one sign factor from $p_i$ ($i=1,2,3$) becomes negative. The cell decomposition plot turns out to be identical to the one before the sign flip, and we also checked that all transition probability matrices remain unchanged.
We note that only the topology of the cell decomposition influences the transition probability matrices. This is why a cell decomposition plot can be viewed as a dual graph. 

For the considered choices of $\tau_1$, $\tau_2$ and $\tau_3$, we calculated all transition probability matrices. We found that, up to permutation of levels and exchange of indices in $p_1$, $p_2$, $p_3$ and $q_1$, $q_2$, $q_3$, there were totally seven types of the transition probability matrices, which we summarized in Table.~I. We  distinguish the transition probability matrices by the number of zeros in their lower triangular part (the number of zeros in the whole matrix is twice this number, since the matrix is always symmetric and all diagonal entries are nonzero). Possible numbers of zeros are: 0, 6, 8, 11, 12, 16. The type 1 (no zeros) contains, in particular, the trivial direct product case, whose  transition probability matrix looks like Eq.~\eqref{eq:P-0}. Besides, we checked that type 3 (eight zeros) contains a direct product of the transition probability matrices of a $2\times2$ LZ model and an $N=2$ $\gamma$-magnet. Among the other five tipes, the type 7 (sixteen zeros) corresponds to the $N=3$ $\gamma$-magnet \cite{gammamagnet}. Types 5 and 6 both have twelve zeros but their distributions of zeros are different,  so these types are not equivalent to each other.
\begin{table}[]\label{table}
\caption{The seven types of transition probability matrices characterized by their numbers of zeros and distributions of zeros. The numbers of zeros in each column are arranged in descending order. }
\smallskip
\begin{tabular}{|c|c|c|}
\hline
Type & Half of the number of zeros  & The number of zeros in columns   \\\hline
1 & 0 & 00000000  \\\hline
2 & 6 &  33111111  \\\hline
3 & 8 & 22222222 \\\hline
4 & 11 & 44332222 \\\hline
5 & 12 & 44333322 \\\hline
6 & 12 & 33333333 \\\hline
7 & 16 & 44444444 \\\hline
\end{tabular}
\end{table}
An example of the transition probability matrices for type 2 (six zeros) is:
\begin{widetext}
\begin{align}
P_{6\,\,\mathrm{zeros}}=\left(\begin{array}{cccccccc}
 p_1 p_2 p_3 & p_2 p_3 q_1 & p_3 q_2 & 0 & p_2 q_3 & 0 & q_2 q_3 & 0 \\
 p_2 p_3 q_1 & p_1 p_2 p_3 & 0 & p_3 q_2 & 0 & p_2 q_3 & 0 & q_2 q_3 \\
 p_3 q_2 & 0 & p_1 p_2 p_3 & p_2 p_3 q_1 & p_1 q_2 q_3 & q_1 q_2 q_3 & p_1 p_2 q_3 & p_2 q_1 q_3 \\
 0 & p_3 q_2 & p_2 p_3 q_1 & p_1 p_2 p_3 & q_1 q_2 q_3 & p_1 q_2 q_3 & p_2 q_1 q_3 & p_1 p_2 q_3 \\
 p_2 q_3 & 0 & p_1 q_2 q_3 & q_1 q_2 q_3 & p_1 p_2 p_3 & p_2 p_3 q_1 & p_1 p_3 q_2 & p_3 q_1 q_2 \\
 0 & p_2 q_3 & q_1 q_2 q_3 & p_1 q_2 q_3 & p_2 p_3 q_1 & p_1 p_2 p_3 & p_3 q_1 q_2 & p_1 p_3 q_2 \\
 q_2 q_3 & 0 & p_1 p_2 q_3 & p_2 q_1 q_3 & p_1 p_3 q_2 & p_3 q_1 q_2 & p_1 p_2 p_3 & p_2 p_3 q_1 \\
 0 & q_2 q_3 & p_2 q_1 q_3 & p_1 p_2 q_3 & p_3 q_1 q_2 & p_1 p_3 q_2 & p_2 p_3 q_1 & p_1 p_2 p_3 \\
\end{array}
\right),
\end{align}
and an example for type 3 (eight zeros) is
\begin{align}
P_{8\,\,\mathrm{zeros}}=
\left(
\begin{array}{cccccccc}
 p_1 p_2 p_3 & p_2 p_3 q_1 & p_1 p_3 q_2 & p_3 q_1 q_2 & p_2 q_3 & 0 & q_2 q_3 & 0 \\
 p_2 p_3 q_1 & p_1 p_2 p_3 & p_3 q_1 q_2 & p_1 p_3 q_2 & 0 & p_2 q_3 & 0 & q_2 q_3 \\
 p_1 p_3 q_2 & p_3 q_1 q_2 & p_1 p_2 p_3 & p_2 p_3 q_1 & q_2 q_3 & 0 & p_2 q_3 & 0 \\
 p_3 q_1 q_2 & p_1 p_3 q_2 & p_2 p_3 q_1 & p_1 p_2 p_3 & 0 & q_2 q_3 & 0 & p_2 q_3 \\
 p_2 q_3 & 0 & q_2 q_3 & 0 & p_1 p_2 p_3 & p_2 p_3 q_1 & p_1 p_3 q_2 & p_3 q_1 q_2 \\
 0 & p_2 q_3 & 0 & q_2 q_3 & p_2 p_3 q_1 & p_1 p_2 p_3 & p_3 q_1 q_2 & p_1 p_3 q_2 \\
 q_2 q_3 & 0 & p_2 q_3 & 0 & p_1 p_3 q_2 & p_3 q_1 q_2 & p_1 p_2 p_3 & p_2 p_3 q_1 \\
 0 & q_2 q_3 & 0 & p_2 q_3 & p_3 q_1 q_2 & p_1 p_3 q_2 & p_2 p_3 q_1 & p_1 p_2 p_3 \\
\end{array}
\right).
\end{align}
\end{widetext}
These matrices show some common features which are also observed in all the 7 types. Namely, all entries are monomials of $p_1$, $p_2$, $p_3$, $q_1$, $q_2$ and $q_3$ with degrees no larger than 3.  All diagonal elements are identically $p_1 p_2 p_3$, which means that the probability to stay in any level is always $p_1 p_2 p_3$.  The transition probabilities between two levels that are directly coupled are always nonzero. For example, consider the transition probabilities from  level 0, given by the entries in the first column of a matrix. Recall that the corresponding vertex 0 is connected to vertices 1, 2 and 4 (see Fig.~\ref{fig:graph-cube}), which means that there are nonzero couplings between level 0 and levels 1, 2 and 4 in the Hamiltonian. We observe that the transition possibilities to these three levels are never zero in all 7 types, whereas transition possibilities to the other four levels can be zero (they are indeed all zero in the 16-zero ($\gamma$-magnet) case).

Here we make a remark related to Ref.~\cite{gammamagnet}. There, the $\gamma$-magnet was presented as an illustration of a phenomenon called dynamic spin localization (DSL) -- for a system of  spins $1/2$. After a linear sweep of the magnetic field, the final state always ends up close to the initial state in the sense that at most one spin flips. This is visualized in the transition probability matrix by the zero entries for the probabilities corresponding to flips of more than one spins. We can thus interpret the number of zeros in a transition probability matrix as a measure of the strength of DSL. Our classification of solutions on cube shows a series of transition probability matrices with numbers of zeros increasing from 0 to that of the $\gamma$-magnet. Thus, the cube model provides a series of Hamiltonians with increasing degrees of DSL, from no DSL (direct product case) to strongest DSL (the $\gamma$-magnet).


\subsection{4-dimensional hypercube}

We further consider a 16-state model whose graph is a 4-dimensional (4d) hypercube, as shown in Fig.~\ref{fig:graph-4dcube}. 
We will show that a 4-dimensional MTLZ family exists on this graph.

Note that inside any (3-dimensional) cube graph inside this 4d hypercube graph, for any two vertices that can be connected by a length-2 path, there are only two such paths in the entire graph. Thus, the analysis for the cube model can be applied as if the cube is an entire graph. We immediately know that all squares (4-loops) in this graph must be non-bipartite, and on any cube graph inside this 4d hypercube graph there is a 3-dimensional family. 
A possible 
directed graph is shown in Fig.~\ref{fig:graph-4dcube}. We label the vertices by binary numbers with 4 digits, from 0000 to 1111. The arrows flow from vertex 1111 to vertex 0000.

\begin{figure}[!htb]
\centering\scalebox{0.5}[0.5]{\includegraphics{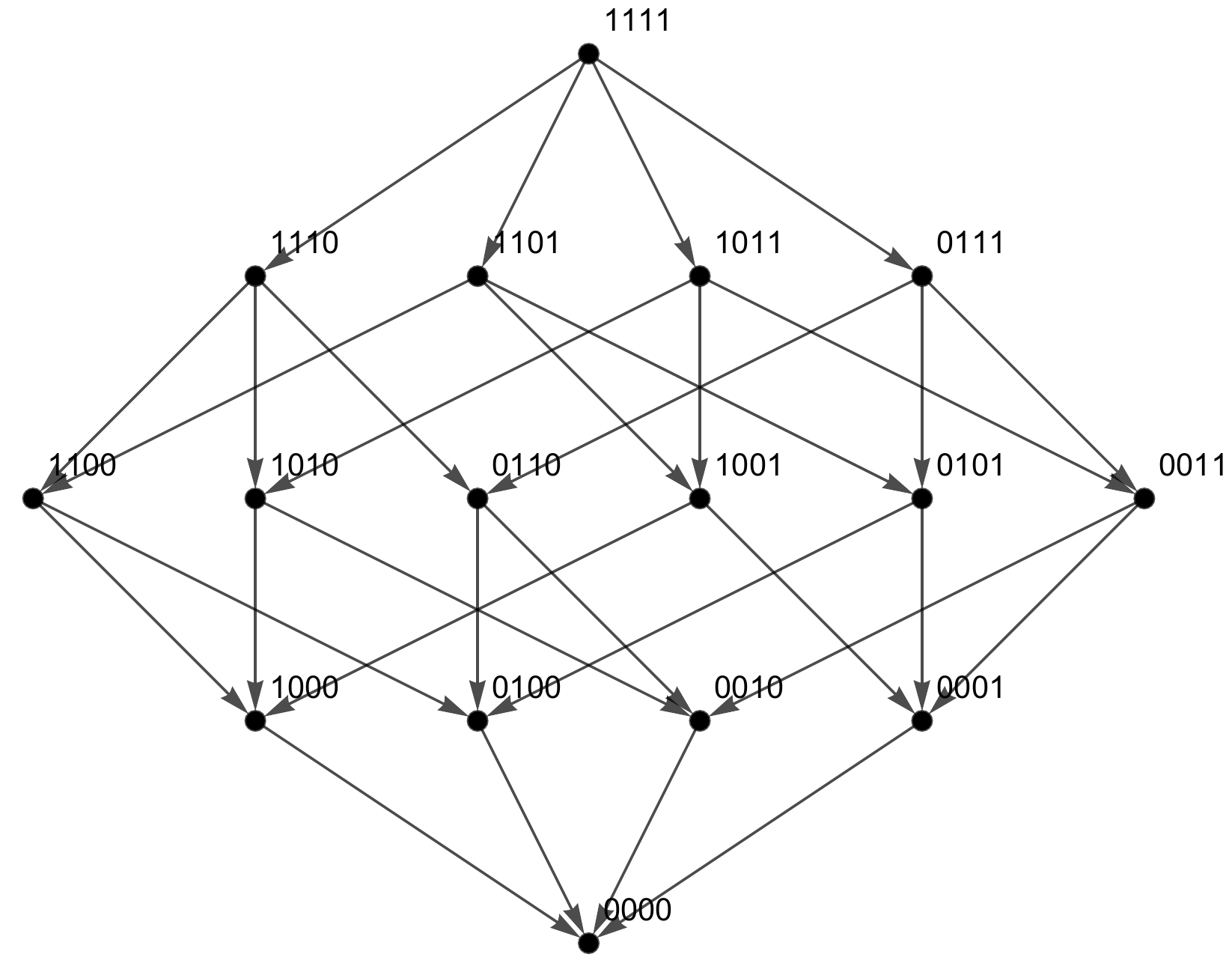}}
\caption{Directed graph of the 4d hypercube model, where every 4-loop has a non-bipartite orientation.}
\label{fig:graph-4dcube}
\end{figure}

We will assume that on this graph 
the dimension of the space of the $\bar A$ forms is 4. We take the $\bar A$ forms on the 4 edges including vertex 0000 to be: $\bar A^{0000,1000}= dx^1$, $\bar A^{0000,0100}= dx^2$, $\bar A^{0000,0010}= dx^3$, $\bar A^{0000,0001}= dx^4$. We also assume that all the 6 squares that contain the vertex 0000 have independent rapidities. We call these rapidities $\vartheta_{12}$,  $\vartheta_{13}$,  $\vartheta_{14}$,  $\vartheta_{23}$, $\vartheta_{24}$, $\vartheta_{34}$, where $\vartheta_{ij}$ corresponds to a square that includes two edges with $dx^i$ and $dx^j$. Correspondingly we denote the 6 hyperbolic tangents as $\tau_{ij}\equiv\tanh\vartheta_{ij}$. We also denote the 6 sign factors for these 6 squares as $p_{ij}$.

Let us now determine the hyperbolic tangent on the square connecting vertices 0011 and 1111. We denote this hyperbolic tangent as  $\tau_{0011,1111}$. There are 2 cubes that include this square: the cube connecting vertices 0010 and 1111, and the cube connecting vertices 0001 and 1111. Both cubes can be used to determine $\tau_{0011,1111}$, and results from both cubes need to be consistent with each other. Let's consider the cube connecting vertices 0010 and 1111. For simplicity, we take the $p$-factor on every square to be 1. According to Eq.~\eqref{eq:tau2}, $\tau_{0011,1111}$ can be expressed as:
\begin{align}
\tau_{0011,1111}=\frac{\tau_{0010,1110}+\tau_{0010,0111}\tau_{0010,1011}}{\sqrt{(1-\tau_{0010,0111}^2)(1-\tau_{0010,1011}^2)}},
\end{align}
and the hyperbolic tangents that appeared in the expression for $\tau_{0011,1111}$ can be expressed in terms of $\tau_{ij}$:
\begin{align}
\tau_{0011,1110}=\frac{\tau_{12}+\tau_{13}\tau_{23}}{\sqrt{(1-\tau_{13}^2)(1-\tau_{23}^2)}},\\
\tau_{0010,0111}=\frac{\tau_{24}+\tau_{23}\tau_{34}}{\sqrt{(1-\tau_{23}^2)(1-\tau_{34}^2)}},\\
\tau_{0010,1011}=\frac{\tau_{14}+\tau_{13}\tau_{34}}{\sqrt{(1-\tau_{13}^2)(1-\tau_{34}^2)}}.
\end{align}
$\tau_{0010,1110}$ then reads in terms of $\tau_{ij}$:
\begin{align}
\tau_{0011,1111}=&[\tau_{12}+\tau_{13}\tau_{23}+\tau_{14}\tau_{24}\nn\\
+&\tau_{34}(\tau_{14}\tau_{23}+\tau_{13}\tau_{24}-\tau_{12}\tau_{34})]/(q_{134}q_{234}),
\end{align}
where we defined
\begin{align}
q_{ijk}=\sqrt{1-\tau_{ij}^2-\tau_{ik}^2-\tau_{jk}^2-2\tau_{ij}\tau_{ik}\tau_{jk}}.
\end{align}
The expression for $\tau_{0011,1111}$ is symmetric in indices 1 and 2, and symmetric in indices 3 and 4. This means that if we use instead the cube connecting vertices 0001 and 1111 to calculate $\tau_{0011,1111}$, the result will be the same, and so the two ways to calculate $\tau_{0011,1111}$ are automatically consistent.

Similarly, we can determine all other hyperbolic tangents for squares including vertex 1111. Thus, all hyperbolic tangents on the 4d hypercube are determined 
by the 6 hyperbolic tangents on squares including vertex 0000. So all $\bar A$ forms are determined by the 4 independent forms $dx^1$, $dx^2$, $dx^3$, $dx^4$. Therefore, there exists a 4-dimensional MTLZ family with 6 independent rapidities on the 4d hypercube graph. A classification of this family should follow the same procedure as in the previous subsection for the cube model, but this classification is expected to be much more complicated, and we will not develop it here.

It is clear now that there must be a rich set of solvable models on the hypercube graphs with dimensions $D>4$. Given the worked out cases with $D=2$, $D=3$ and $D=4$, we can speculate that for $D>4$ the highest
dimension of the MTLZ family is also $D$, i.e., it contains $D$ independent Hamiltonians, and the number of independent rapidities is $D(D-1)/2$. This  family contains the trivial model of $D$ independent spins, which is obtained if we set all rapidities to zero. We leave this conjecture without proof, as well as leave  the question open about the existence of other families for hypercubes with $D\ge4$. 

\section{Fan}


In addition to cube, we explored connectivity graphs with other topology for possibilities to satisfy the integrability conditions. We found that we can satisfy the integrability conditions for the ``fan'' model that we show in Fig.~\ref{fig:graph-fan}.
This model contains $m+2$ vertices, with $m$ vertices $\{a_{1}, \ldots, a_{m}\}$ all connecting to two other vertices $b_{1}$ and $b_{2}$ but not connecting among themselves. Later we will refer to these two types of vertices as $a$-vertices and $b$-vertices, respectively. 
We found that this model corresponds to a $2$-dimensional family that has been already studied by us in \cite{large-class}.

\begin{figure}[!htb]
 (a)\scalebox{0.45}[0.45]{\includegraphics{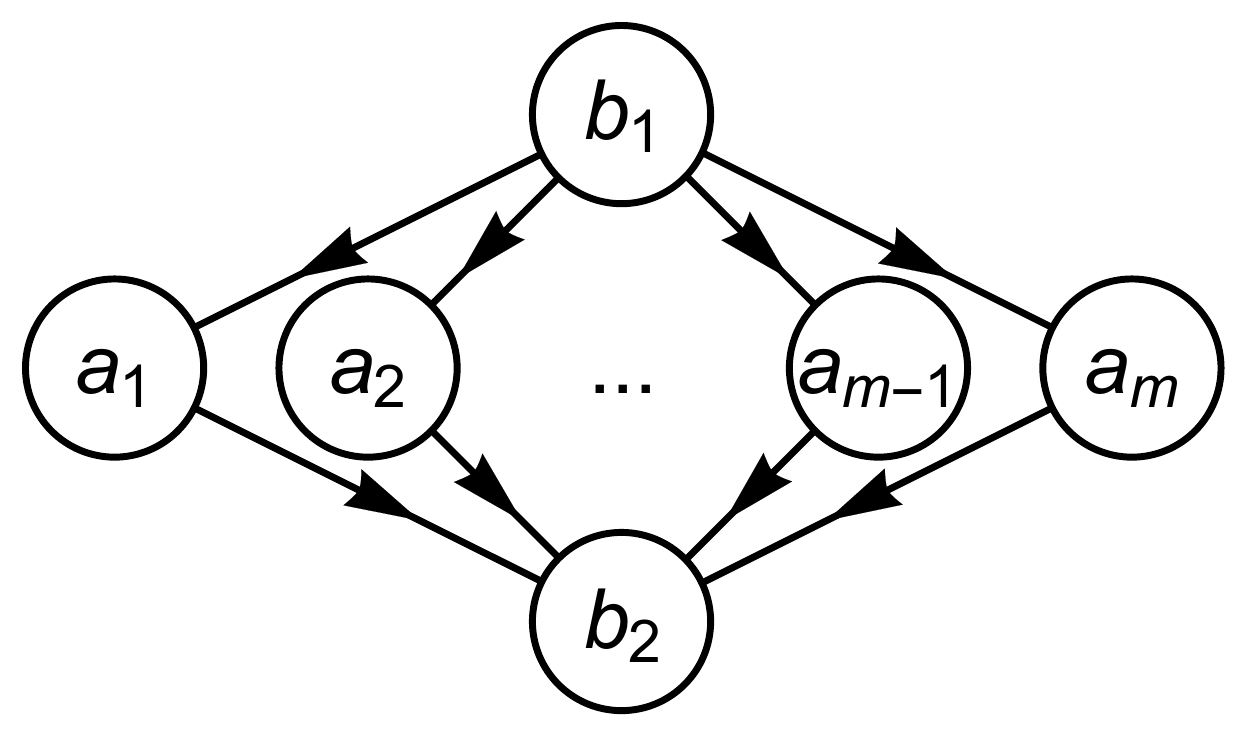}}
 (b)\scalebox{0.45}[0.45]{\includegraphics{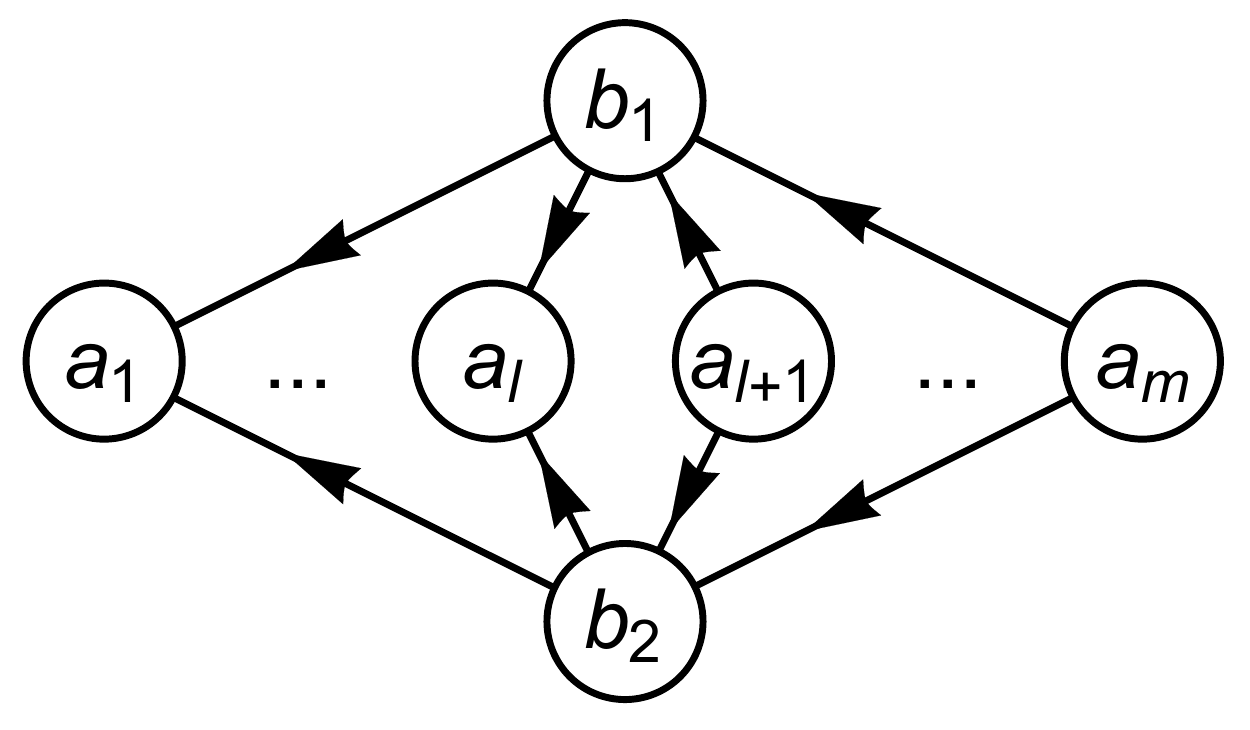}}
\caption{The graph of the ``fan" model: $m$ states interact only with two other states. (a) Directed graph of the type-I orientation. (b) Directed graph of the type-II orientation.}
\label{fig:graph-fan}
\end{figure}


Let's identify all allowed orientations on the fan graph. We introduce a convenient notation $\alpha_{j} = \{b_{1}, a_{j}\}$ and $\beta_{j} = \{b_{2}, a_{j}\}$ for $j = 1, \ldots, m$. 
Suppose that one of the $a$-vertices, say $a_{1}$, is intermediate. By considering $4$-loops $(\alpha_{1}, \beta_{1}, \beta_{k}, \alpha_{k})$ we see that all $a_{k}$ are intermediate and if $b_{1}$ is a source/sink in one of the loop-generated graphs it is a source/sink in all others. This implies that $b_{1}$ and $b_{2}$ is a source and a sink, respectively, or vice versa; in other words there is a unique orientation of this kind up to a permutation of $b_{1}$ and $b_{2}$. Suppose now that $a_{1}$ is not intermediate, so, say it is a source. By the same argument as the one just above we see that any other $a_{j}$ is either a source or a sink. This implies that up to a permutation of $a$-vertices there are $m$ possible orientations of this kind, labeled by $1 \le l \le m$, with $a_{1}, \ldots, a_{l}$ and $a_{l+1}, \ldots, a_{m}$ being sinks and sources, respectively. The described orientations are referred to as type-I and type-II orientations, 
respectively, and they are shown in Fig.~\ref{fig:graph-fan}(a) and (b), respectively.

We are now in a position to identify all solutions of Eq.~(\ref{cycle-property-bar-A}) for the graphs of the type of $\hat{\Gamma}^{ab}$, by applying the classification of solutions for $4$-loop generated graphs. 
To that end we note all $4$-loops of the considered graphs are parameterized by pairs of distinct $a$-vertices, i.e., by ordered pairs $(j, k)$ with $1 \le j < k \le m$ that represent the loops $(\alpha_{j}, \beta_{j}, \beta_{k}, \alpha_{k})$. Denoting $\bar A^{j} = (\bar A^{\alpha_{j}}, \bar A^{\beta_{j}})$, we apply 
the properties of the local solutions to obtain
\begin{align}
\label{solution-via-U} \bar A^{j} = U_{jk} \bar A^{k}, \;\;\; U_{jl} = U_{jk} U_{kl}, \;\;\; \forall \, m \ge j > k > l \ge 1,
\end{align}
with $U_{jk}$ being $2 \times 2$ matrices which are orthogonal or pseudoorthogonal, and the second set of equalities are the consistency conditions. We can eliminate all consistency conditions by parameterizing a solution by a set $(U_{m, m-1}, \ldots, U_{21})$ of matrices with the others explicitly expressed by
\begin{eqnarray}
\label{solution-via-U-2} U_{jk} = U_{j, j-1} \ldots U_{k+1, k}, \;\;\; {\rm for} \;\; j-k > 1.
\end{eqnarray}
Since all $U_{jk}$ matrices belong to the orthogonal or pseudoorthogonal  group, implementation of Eqs.~(\ref{solution-via-U-2}) is an easy task.

In the rest of this section we will view the fan graph 
as an entire graph, and find an explicit solution of Eq.~(\ref{constr-3-LA-bar-A}). We start with demonstrating that type-I orientation shown in Fig.~\ref{fig:graph-fan}(a) does not support non-trivial solutions. Indeed, for a pair of $a$-vertices, say $a_{j}$ and $a_{k}$, the r.h.s. of Eq.~(\ref{constr-3-LA-bar-A}) has two terms, and, combined with Eq.~(\ref{A-O-determinants}), we derive
\begin{eqnarray}
\label{constr-3-K-2m} &\sqrt{\gamma^{\alpha_{j}} \gamma^{\alpha_{k}}} \bar A^{\alpha_{j}} \wedge \bar A^{\alpha_{k}} + \sqrt{\gamma^{\beta_{j}} \gamma^{\beta_{k}}} \bar A^{\beta_{j}} \wedge \bar A^{\beta_{k}} \nn\\
&= (\sqrt{\gamma^{\alpha_{j}} \gamma^{\alpha_{k}}} + r_{jk} \sqrt{\gamma^{\beta_{j}} \gamma^{\beta_{k}}}) \bar A^{\alpha_{j}} \wedge \bar A^{\alpha_{k}} = 0,
\end{eqnarray}
and to have a nontrivial solution we should set all sign factors $r_{jk} = -1$. This leads to  $\bar A^{\alpha_{j}} \wedge \bar A^{\alpha_{k}}=- \bar A^{\beta_{j}} \wedge \bar A^{\beta_{k}}$ for any $j$ and $k$. 
For type-I orientation all 4-loops are non-bipartite, so (due to Eqs.~(\ref{A-LO-transf-correspond-3}) and (\ref{A-LO-determinants-tilde})) we have $\bar A^{\alpha_{j}} \wedge \bar A^{\beta_{j}} =- \bar A^{\alpha_{k}} \wedge \bar A^{\beta_{k}}$ for all distinct pairs. This leads to contradictions for $m\ge 3$ (when there are at least three $a_j$ vertices), since $\bar A^{\alpha_{1}} \wedge \bar A^{\beta_{1}} =- \bar A^{\alpha_{2}} \wedge \bar A^{\beta_{2}}$ and $\bar A^{\alpha_{1}} \wedge \bar A^{\beta_{1}} =- \bar A^{\alpha_{3}} \wedge \bar A^{\beta_{3}}$ togeither would imply $\bar A^{\alpha_{2}} \wedge \bar A^{\beta_{2}} = \bar A^{\alpha_{3}} \wedge \bar A^{\beta_{3}}$. So there are no non-trivial solutions.

We now apply the same kind of analysis to the type-II orientation case shown in Fig.~\ref{fig:graph-fan}(b). 
Without loss of generality, we set all vertices $a_{j}$ with $1 \le j \le l$ and $l+1 \le j \le m$ to be sinks and sources, respectively, as shown in Fig.~\ref{fig:graph-fan}(b). Since all graphs produced by $4$-loops that include $a_{k}$ and $a_{j}$ with $1 \le k < j \le l$ (sink region) or $l+1 \le k < j \le m$ (source region), respectively have bipartite orientation, the matrices $U_{jk}$ in this range are of the type given by Eq.~(\ref{A-LO-transf}), so that Eq.~(\ref{constr-3-K-2m}), where, due to the chosen notation (compare Eq.~(\ref{A-LO-transf}) with Eq.~(\ref{A-LO-transf-LO})), $r_{jk}$ should be replaced by $\tilde{r}_{jk}$, implies $\tilde{r}_{jk} = -1$. According to Eq.~(\ref{A-LO-transf-LO-3}), we then have $r_{jk} = 1$ in the aforementioned range, so that
\begin{align}
\label{constr-3-K-2m-3}  &\bar A^{\alpha_{j}} \wedge \bar A^{\beta_{j}} =\bar A^{\alpha_{k}} \wedge\bar A^{\beta_{k}}, \;\;\; {\rm for} \;\; 1 \le k < j \le l,\nn\\
&\bar A^{\alpha_{j}} \wedge\bar A^{\beta_{j}} =\bar A^{\alpha_{k}} \wedge\bar A^{\beta_{k}}, \;\;\; {\rm for}\;\; l+1 \le k\le j \le m.
\end{align}
A similar consideration for the $4$-loop that includes $a_{l}$ and $a_{l+1}$, which has a non-bipartite orientation yields $\bar A^{\alpha_{l}} \wedge \bar A^{\beta_{l}} = \bar A^{\alpha_{l+1}} \wedge \bar A^{\beta_{l+1}}$, and Eq.~(\ref{constr-3-K-2m-3}) can be extended by
\begin{align}
\label{constr-3-K-2m-4}  \bar A^{\alpha_{j}} \wedge\bar A^{\beta_{j}} = -\bar A^{\alpha_{k}} \wedge\bar  A^{\beta_{k}}, \;\;\; {\rm for} \;\; 1 \le k \le l < j \le m,
\end{align}
so that Eq.~(\ref{constr-3-LA-bar-A}) takes a form
\begin{align}
\label{constr-3-K-2m-5} \bar A^{\alpha_{1}} \wedge \bar A^{\beta_{1}} \left(\sum_{j=1}^{l} \sqrt{\gamma^{\alpha_{j}} \gamma^{\beta_{j}}} - \sum_{j=l+1}^{m} \sqrt{\gamma^{\alpha_{j}} \gamma^{\beta_{j}}}\right) = 0.
\end{align}

More careful analysis of Eq.~(\ref{constr-3-K-2m}), i.e., analyzing it for any pair of $a$-vertices out of three, say $a_{j}$, $a_{k}$, and $a_{q}$ shows that the equality holds for any $4$-loop, if and only if $\sqrt{\gamma^{\alpha_{j}}} = \sqrt{\gamma^{\beta_{j}}}$, for all $1 \le j \le m$. This combined with Eq.~(\ref{constr-3-K-2m-5}) finally yields
\begin{align}
\label{constr-3-K-2m-6}  \sum_{j=1}^{l} \gamma^{\alpha_{j}} - \sum_{j=l+1}^{m} \gamma^{\alpha_{j}} = 0, \;\;\; \gamma^{\beta_{j}} = \gamma^{\alpha_{j}}, \;\;\; {\rm for} \;\; 1 \le j \le m.
\end{align}
The overall sign factors $p_{j+1, j}$, for $j = 1, \ldots, m-1$ can be chosen in an arbitrary way.

Note that, according to the way Eq.~(\ref{A-LO-transf}) and Eq.~(\ref{A-LO-transf-LO}) are represented in terms of ordering of the edges, we have
\begin{align}
\label{U-explicit} 
&U_{l+1, l} = 
p_{l+1, l} U(\vartheta_{l+1, l}) \sigma=p_{l+1, l}\sigma U(\vartheta_{l+1, l}) , 
\nn\\ 
&U_{j+1, j} = p_{j+1, j} U(\vartheta_{j+1, j}), \;\;\; {\rm for} \;\; 1 \le j \le m-1, \;\; {\rm and} \;\; j \ne l,
\end{align}
where $\sigma = \sigma_{x}$ is the $2 \times 2$ permutation matrix, and $U(\vartheta)$ has a form of Eq.~(\ref{A-LO-transf}) with $p = r = 1$. Also note that for Eq.~(\ref{constr-3-K-2m-6}) to be satisfied we should have $l \ne 1, m$.

Summarizing, for a fan graph, 
a solution of the system of equations that represents the integrability conditions for a linear multistate LZ family is completely parameterized by the following data: (i) An integer number $l$ with $1 < l < m$, (ii) a set $(p_{j+1, j} = \pm 1 \, | \, j = 1, \ldots, m-1)$ of sign factors, (iii) a set $(\vartheta_{j+1, j} \in \mathbb{R} \, | \, j = 1, \ldots, m-1)$ of rapidities, and (iv) a set $(\gamma^{j} >0 \, | \, j = 1, \ldots, m)$ of strictly positive LZ parameters that satisfy the constraint
\begin{eqnarray}
\label{constr-for-gamma}  \sum_{j=1}^{l} \gamma^{j} - \sum_{j=l+1}^{m} \gamma^{j} = 0,
\end{eqnarray}
so that, denoting $\bar{A}^{j} = (\bar{A}^{\alpha_{j}}, \bar{A}^{\beta_{j}})$, we have
\begin{align}
\label{A-via-data} 
&\bar{A}^{j} = p_{j,1} U(\vartheta_{j, 1})\bar{A}^{1}, \;\;\; {\rm for} \;\; 1 \le j \le l,\nn\\
&\bar{A}^{j} = p_{j,1} \sigma U(\vartheta_{j, 1})\bar{A}^{1}, \;\;\; {\rm for}  \;\;l+1 \le j \le m; \nonumber \\ 
&p_{jk} = \prod_{q=k}^{j-1} p_{q+1, q}, \;\;\; \vartheta_{jk} = \sum_{q=k}^{j-1} \vartheta_{q+1, q}, \;\;\; {\rm for} \;\; 1 \le k < j \le m; \nn\\
&\gamma^{\alpha_{j}} = \gamma^{\beta_{j}} = \gamma^{j}, \;\;\; {\rm for} \;\; 1 \le j \le m; \nonumber \\ 
&s^{a_{j}b_{1}} = s^{a_{j}b_{2}} = 1, \;\;\; {\rm for} \;\; 1 \le j \le l,\nn\\ 
&s^{a_{j}b_{1}} = s^{a_{j}b_{2}} = -1,\;\;\; l+1 \le j \le m.
\end{align}


\section{Graphs that do not sustain integrable families}


\begin{figure}[!htb]
\centering \scalebox{0.3}[0.3]{\includegraphics{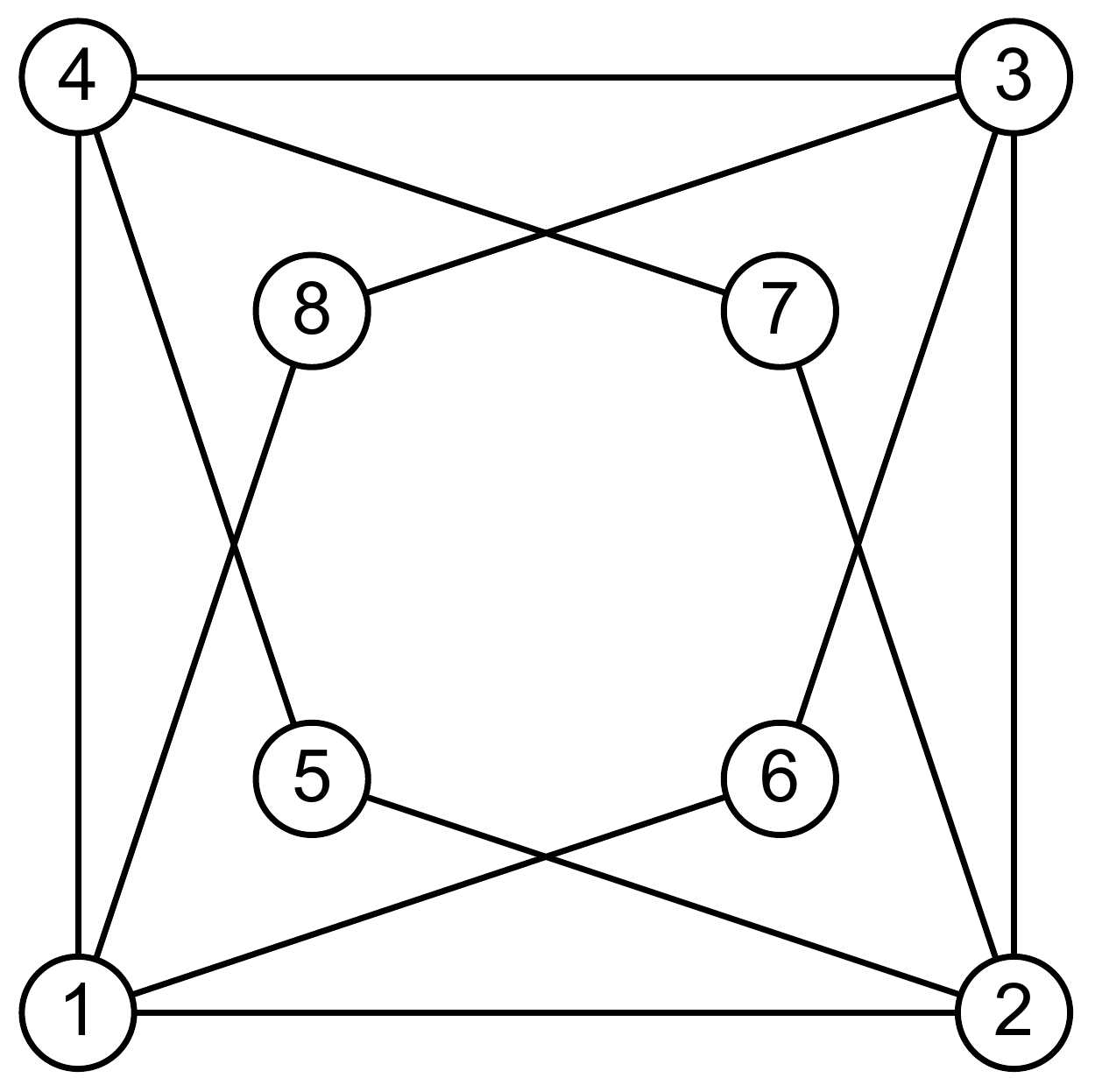}}
\caption{The ``double-fan'' graph. 
}
\label{fig:double-fan}
\end{figure}

\begin{figure}[!htb]
 \scalebox{0.35}[0.35]{\includegraphics{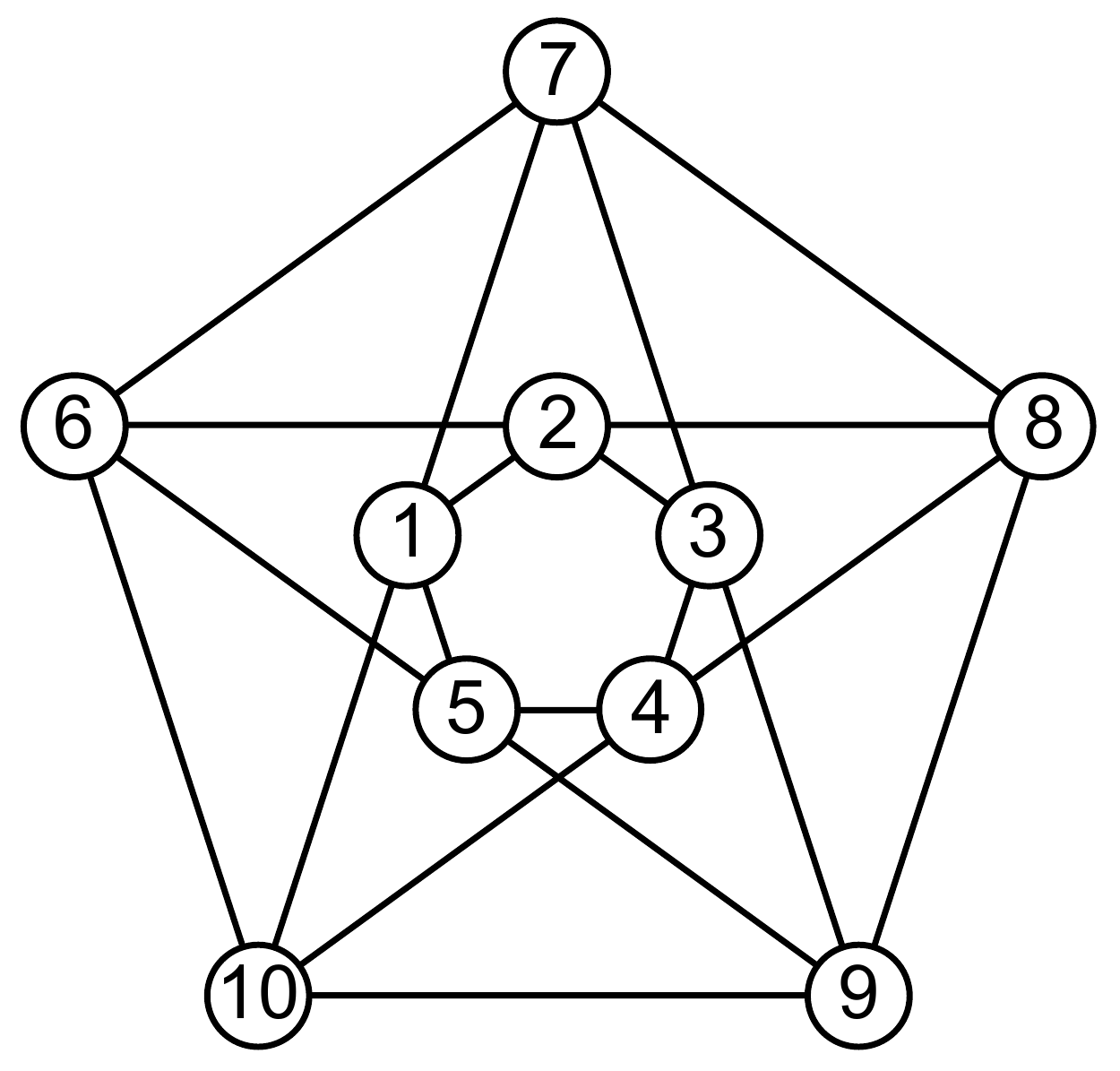}}
\caption{The ``double-pentagon'' graph.}
\label{fig:star-fig}
\end{figure}

\begin{figure}[!htb]
\scalebox{0.35}[0.35]{\includegraphics{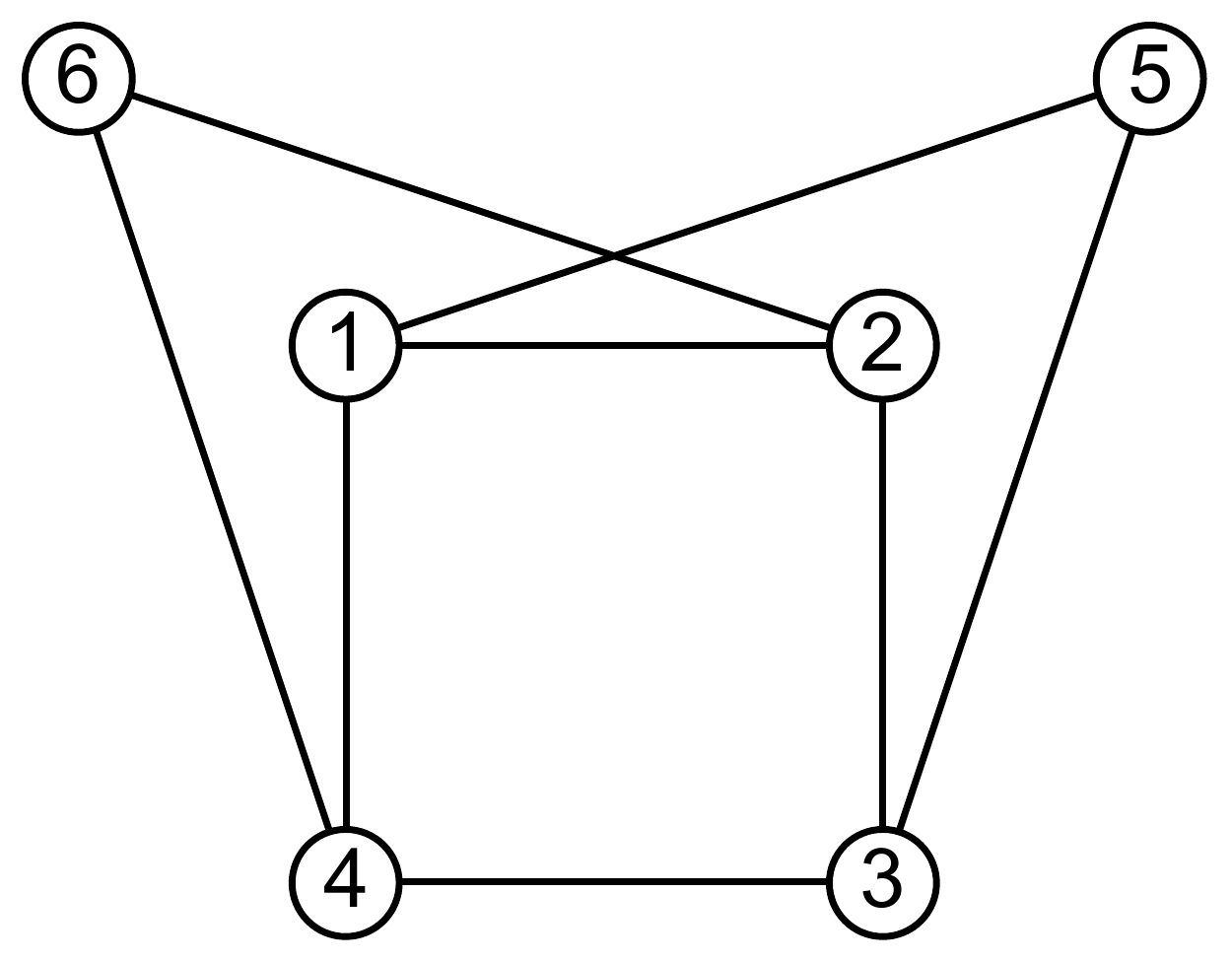}}
\caption{The ``square with ears'' graph.}
\label{fig:graph-ears}
\end{figure}

\begin{figure}[!htb]
\scalebox{0.3}[0.3]{\includegraphics{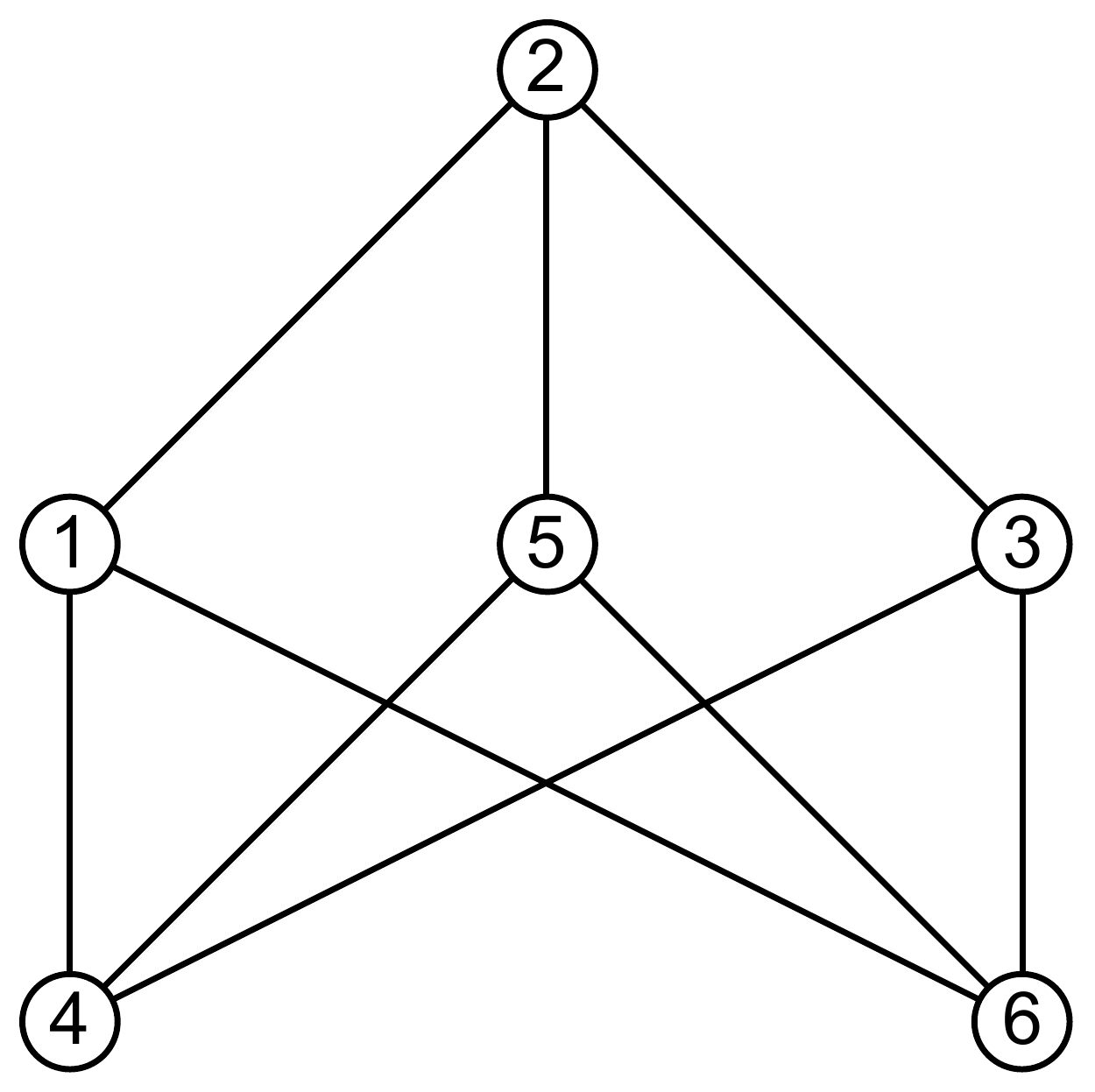}}
\caption{The ``M\"{o}bius ladder'' graph. 
}
\label{fig:graph-Mobius}
\end{figure}

\begin{figure}[!htb]
\centering \scalebox{0.35}[0.35]{\includegraphics{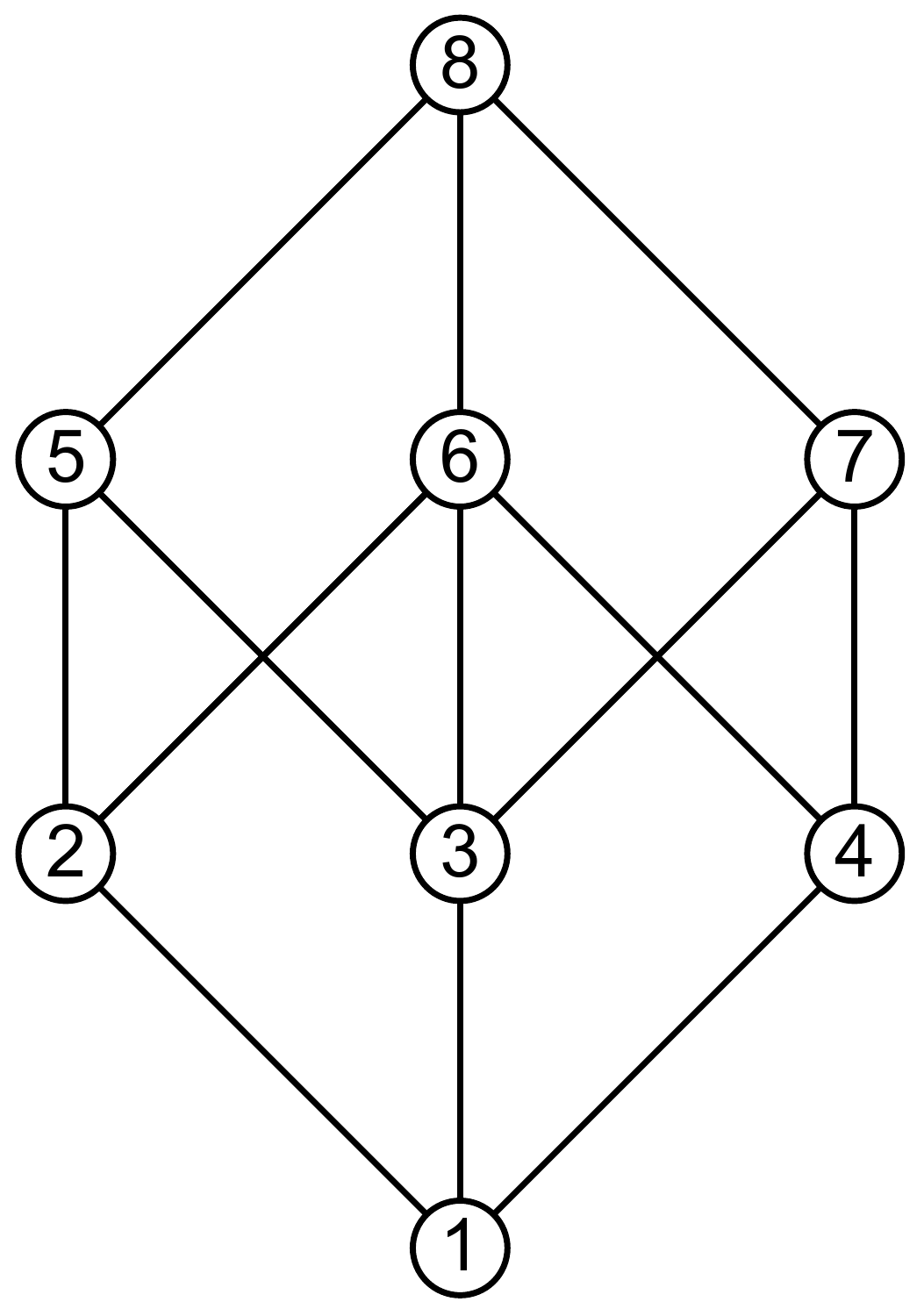}}
\caption{The ``cube$+1$'' graph.}
\label{fig:cube+1diagonal}
\end{figure}

\begin{figure}[!htb]
\centering \scalebox{0.35}[0.35]{\includegraphics{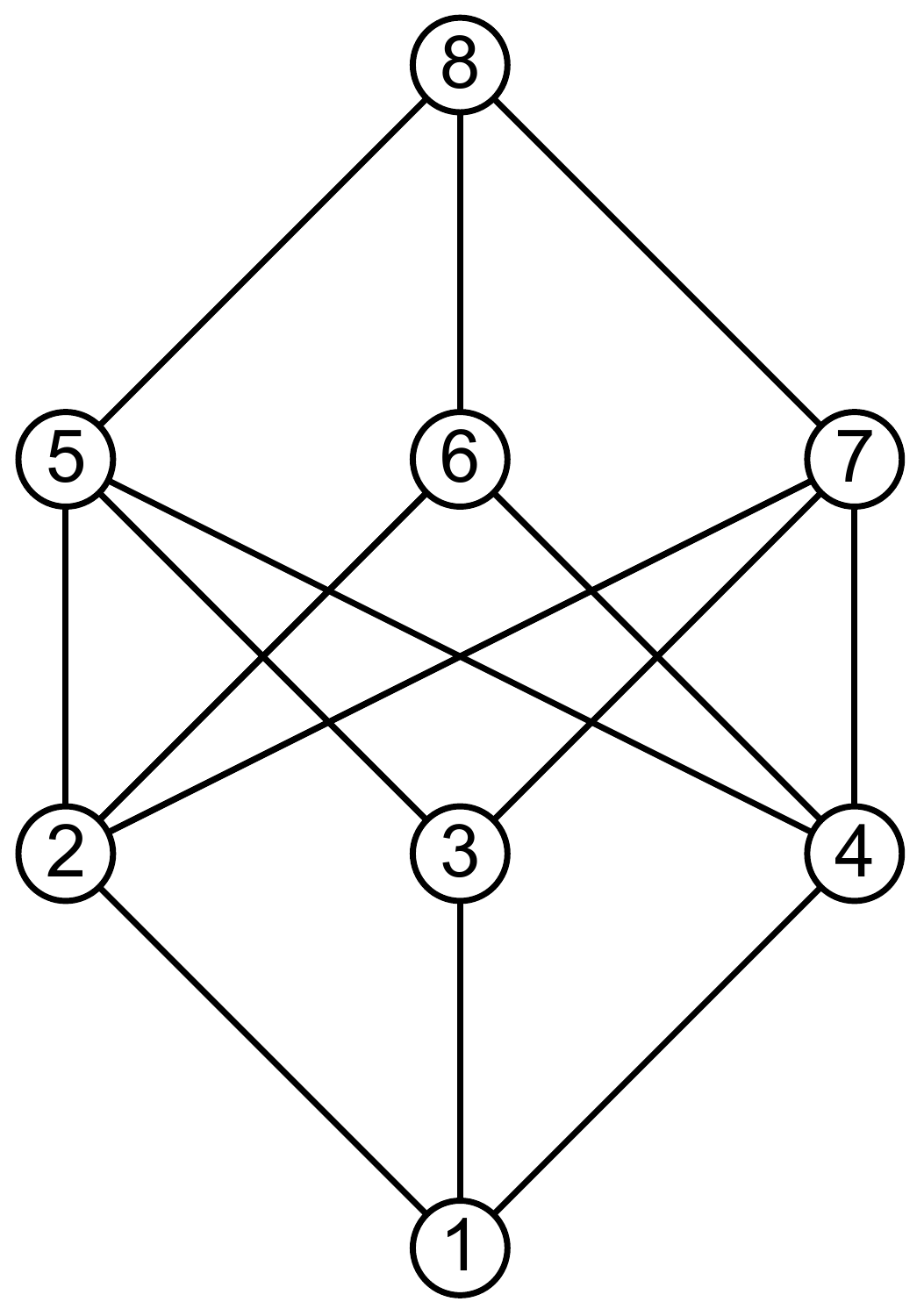}}
\caption{The ``cube$+2$'' graph.}
\label{fig:cube+2diagonal}
\end{figure}

\begin{figure}[!htb]
\centering \scalebox{0.35}[0.35]{\includegraphics{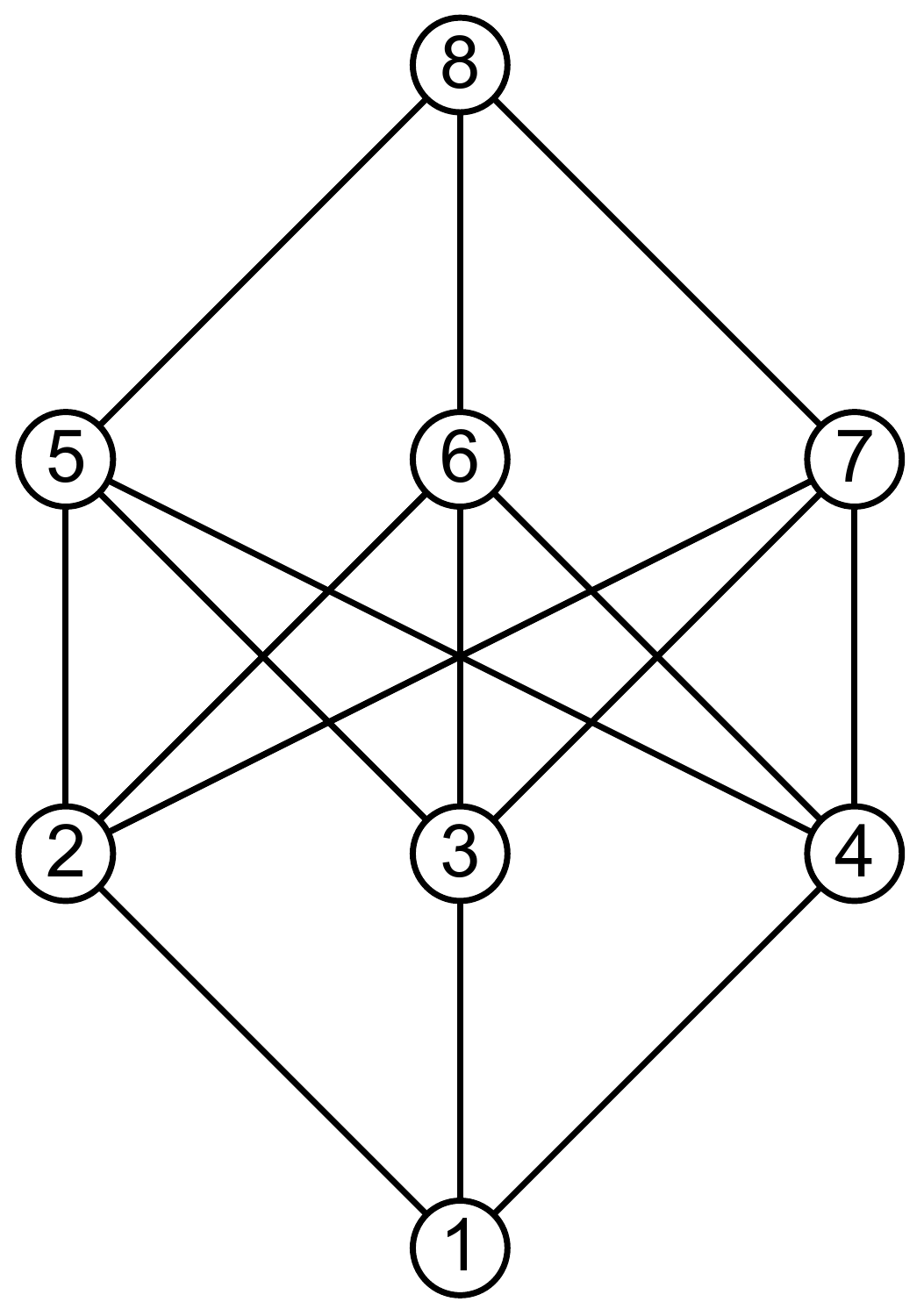}}
\caption{The ``cube$+3$'' graph.}
\label{fig:cube+3diagonal}
\end{figure}

Finally, we would like to  mention also the types of graphs for which we checked that the integrability conditions cannot be satisfied. The first such a graph is shown in Fig.~\ref{fig:double-fan}. We called it a ``double-fan'', since it can be viewed as two fans intertwining with each other. Note that this graph can be obtained if we replace two edges of a cube by two diagonal links.

The analysis for this graph goes as follows. First, we assume that the 4-loop 1234 is non-bipartite, say vertex 1 is a source, 3 is a sink, and 2, 4 are intermediate. Then the fan with $b$-vertices 1 and 3 is of type-I. Consider the length-2 paths condition (Eq.~\eqref{constr-3-LA-bar-A}) between vertices 2 and 6, we see that $\bar A^{12}\wedge \bar A^{16}=-\bar A^{23}\wedge \bar A^{36}$, and since loop 1236 is non-bipartite, we have $\bar A^{12}\wedge \bar A^{23}=-\bar A^{16}\wedge \bar A^{36}$. Similarly, condition \eqref{constr-3-LA-bar-A} between vertices 2 and 8 gives $\bar A^{12}\wedge \bar A^{23}=-\bar A^{18}\wedge \bar A^{38}$, and condition \eqref{constr-3-LA-bar-A} between vertices 6 and 8 gives $\bar A^{16}\wedge \bar A^{36}=-\bar A^{18}\wedge \bar A^{38}$. But these three equations are contradictory, so the loop 1234 cannot be non-bipartite. Let's then assume that this loop is bipartite, for which we can choose the sources to be 1 and 3 without loss of generality. Now consider the loops 1236 and 1436. Condition \eqref{constr-3-LA-bar-A} on vertices 2 and 6 gives $\bar A^{12}\wedge \bar A^{16}=-\bar A^{23}\wedge \bar A^{36}$, and condition \eqref{constr-3-LA-bar-A} on vertices 4 and 6 gives $\bar A^{14}\wedge \bar A^{16}=-\bar A^{34}\wedge \bar A^{36}$. Vertex 6 can be a source or a sink, but in either case we will have $\bar A^{12}\wedge \bar A^{23}=\bar A^{14}\wedge \bar A^{34}$. If we consider the loops 2145 and 2345, the same argument gives $\bar A^{12}\wedge \bar A^{14}=\bar A^{23}\wedge \bar A^{34}$. However, since  loop 1234 is bipartite, $\bar A^{12}\wedge \bar A^{23}=r\bar A^{14}\wedge \bar A^{34}$ will give $\bar A^{12}\wedge \bar A^{14}=-r\bar A^{23}\wedge \bar A^{34}$, so we still get a contradiction. Therefore, the ``double-fan'' graph does not support a solution.

In principle, our analysis does not exclude the possibility that MTLZ families can be constructed on graphs that contain longer than $4$-edge loops. An example of such a candidate is shown in Fig.~\ref{fig:star-fig}, which we call the ``double-pentagon'' graph. However, our analysis shows that it does not sustain a solution. 
Indeed, in Fig.~\ref{fig:star-fig}, let's consider the fan graph made by vertices 1, 2, 3, 6, 7 and 8. We will call it ``fan (2,7)'', since its $b$-vertices are 2 and 7. This fan can be viewed as being composed of three 4-loops:  the 4-loop 1237 belongs solely to this fan, and the 4-loop 1267 and the 4-loop 2378 are shared by the neighboring two fans. According to arguments of the previous section, this fan is of either type-I or type-II. Let's first consider the case when the fan is of type-I, so all the 4-loops of this fan are non-bipartite. We will try to get relations of the $\bar A$ forms in loop 2378. To do so we first note that vertices 1 and 3 are connected by only two length-2 paths, so $\bar A^{12} \wedge\bar A^{23} = -\bar A^{17} \wedge\bar  A^{37}$. Since loop 1237 is non-bipartite, we have $\bar A^{12} \wedge\bar A^{17} =-\bar A^{23} \wedge\bar  A^{37}$. In loop 1287, similarly we get $\bar A^{12} \wedge\bar A^{17} =-\bar A^{28} \wedge\bar  A^{78}$. These two equations togeither lead to $\bar A^{23} \wedge\bar  A^{37}=\bar A^{28} \wedge\bar  A^{78}$. Since loop 2378 is non-bipartite, we further get $\bar A^{23} \wedge\bar  A^{28}=\bar A^{37} \wedge\bar  A^{78}$. If fan (2,7) is of type-II, we can follow the same steps to obtain relations for  $\bar A$ forms in loop 2378. There are three cases:\\
Case 1. When fan (2,7) is of type-I:
\begin{align}
\bar A^{23} \wedge\bar  A^{37}=\bar A^{28} \wedge\bar  A^{78},\quad \bar A^{23} \wedge\bar  A^{28}=\bar A^{37} \wedge\bar  A^{78}.
\end{align}
Case 2. When fan (2,7) is of type-II and loop 2378 is non-bipartite:
\begin{align}
\bar A^{23} \wedge\bar  A^{37}=-\bar A^{28} \wedge\bar  A^{78},\quad \bar A^{23} \wedge\bar  A^{28}=-\bar A^{37} \wedge\bar  A^{78}.
\end{align}
Case 3. When fan (2,7) is of type-II and loop 2378 is bipartite:
\begin{align}
\bar A^{23} \wedge\bar  A^{37}=\bar A^{28} \wedge\bar  A^{78},\quad \bar A^{23} \wedge\bar  A^{28}=-\bar A^{37} \wedge\bar  A^{78}.
\end{align}
We can perform exactly the same argument for all other fans inside the double-pentagon graph, especially for the fan made by vertices 2, 3, 4, 7, 8, 9 (denote it as ``fan (3,8)''):\\
Case 1. When fan (3,8) is of type-I:
\begin{align}
\bar A^{23} \wedge\bar  A^{37}=\bar A^{28} \wedge\bar  A^{78},\quad \bar A^{23} \wedge\bar  A^{28}=\bar A^{37} \wedge\bar  A^{78}.
\end{align}
Case 2. When fan (3,8) is of type-II and loop 2378 is non-bipartite:
\begin{align}
\bar A^{23} \wedge\bar  A^{37}=-\bar A^{28} \wedge\bar  A^{78},\quad \bar A^{23} \wedge\bar  A^{28}=-\bar A^{37} \wedge\bar  A^{78}.
\end{align}
Case 3. When fan (3,8) is of type-II and loop 2378 is bipartite:
\begin{align}
\bar A^{23} \wedge\bar  A^{37}=-\bar A^{28} \wedge\bar  A^{78},\quad \bar A^{23} \wedge\bar  A^{28}=\bar A^{37} \wedge\bar  A^{78}.
\end{align}
We see that the two sets of relations for  $\bar A$ forms are consistent only when both fans (2,7) and  (3,8) are of type-I, or when both  fans (2,7) and (3,8) are of type-II and loop 2378 is non-bipartite. However, neither of these two situations is possible. If  fan (2,7) is of type-I, then vertex 2 is either a source or a sink, so fan (3,8) must be of type-II. Conversely, if fan (2,7) is of type-II and loop 2378 is non-bipartite, then vertex  2 is intermediate in loop 2378, so fan (3,8) must be of type-I. So no solutions are possible on the double-pentagon graph. Note that for any other graph which has the same structure as Fig.~\ref{fig:star-fig} but with the two pentagons replaced by two polygons with any larger number of edges (e.g. a ``double-hexagon'' graph), the same argument can be applied to show that it also does not support solutions. 

We also considered several other graphs. 
For the ``square with ears" graph in Fig.~\ref{fig:graph-ears}, the ``M\"{o}bius ladder'' graph 
in Fig.~\ref{fig:graph-Mobius}, and the ``cube$+1$'' graph in Fig.~\ref{fig:cube+1diagonal} which is constructed by connecting one diagonal on the cube graph, 
we analyzed all possible orientations and found that trying to satisfy all integrability conditions always lead to contradictions. 
We also checked certain orientations of the ``cube$+2$'' graph in Fig.~\ref{fig:cube+2diagonal} and the ``cube$+3$'' graph in Fig.~\ref{fig:cube+3diagonal} constructed by connecting two or three diagonals on the cube graph, and did not find solutions but we did not pursue the rigorous no-go proof in these two cases.

The numerous ``no-go" examples suggest that the hypercube, the fan family, as well as their various deformed direct products \cite{large-class}, are the only  independent MTLZ families that are
possible.  We leave such conjectures for future studies.

\section{Discussion}


The multitime Landau-Zener model (\ref{linear-family}), when it is supplemented with integrability conditions (\ref{linear-family-2})-(\ref{linear-family-4}), defines a set of high order linear ordinary differential equations, whose solutions can be well described analytically and classified.  The model (\ref{linear-family}) has one  irregular singular point at $t=\infty$ as the parabolic cylinder equation, and hence shares similar analytical properties with it. Therefore, it is convenient to think about  the model (\ref{linear-family})  as defining a new special function that generalizes the 2nd order parabolic cylinder function.
There are several other properties of the MTLZ model that characterize it as defining a physically useful special function:

(i) It describes quantum mechanical evolution that represents a broad physically interesting type of processes. Importantly, the MTLZ model defines not a single model but rather a large class of solvable equations. For most of the allowed values of parameters, physical meaning of the Hamiltonian, e.g., the interpretation in terms of interacting spins, is yet to be found. However, the analytical description of the time-dependent evolution can be developed in advance, as it happened with many commonly used special functions.

(ii) As for many standard special functions, it is possible to connect asymptotic behavior of our solutions at $t\rar \pm \infty$. At least several other properties, such as the presence of a specific number of exact eigenvalue crossing points, are possible to prove analytically. It is also likely that a solution for arbitrary time can be found in terms of contour integrals, as it was shown for multistate LZ models that are related to the Gaudin magnet family \cite{yuzbashyan}.

(iii) The  Hamiltonian (\ref{linear-family}) is sufficiently simple, so that one can use it as a compact definition of the set of free parameters.

(iv) The simplicity of an analytical solution usually matters for applications in physics. The  transition probabilities in the models from the MTLZ families are expressed in terms of elementary functions of the model's
parameters  \cite{large-class}. In this sense, behavior of our systems are often much easier to understand than, e.g., physics of stationary models  that are solvable by the Bethe ansatz.


By no means the MTLZ family exhausts the class of solvable multistate LZ models. A simple counterexample is the Demkov-Osherov model that belongs to a family whose all other independent Hamiltonians depend nonlinearly on time-like variables \cite{yuzbashyan}.  
The present article shows rather that by restricting the multi-time dependence of the Hamiltonians to relatively simple functions of all time variables, it is possible to fully classify and achieve a very detailed understanding of the scattering matrix for any given number of interacting states. The program that we described can be, in principle, fully automated using mathematical software for symbolic calculations. 

Interestingly, even after achieving a complete classification up to some finite $N$ of interacting states, it  remains hard to identify the cases with presently useful physical interpretation.  Thus, even for a square graph, the physically interesting $\gamma$-magnet Hamiltonian appeared at a nontrivial value of the rapidity variable. We did not explore how to separate such particularly interesting models from the rest of the family.
 Historically, most of the commonly known special functions were studied for the possibility to understand the equations that had these functions as solutions. Only later found this many applications in physics. Therefore, we suggest that the new families of integrable models  must be studied for their own sake, as they define new special functions that will be needed  for the future research on strongly interacting quantum systems.

\section*{Acknowledgements}
This work was supported by the U.S. Department of Energy, Office of Science, Basic Energy
Sciences, Materials Sciences and Engineering Division, Condensed Matter Theory Program (V.Y.C. and N.A.S.), and by the J. Michael Kosterlitz Postdoctoral Fellowship at Brown University (C.S.).

Authors made equal contributions to this article.
\appendix

\section{Properties of Good MTLZ Families of Integrable Hamiltonians}
\label{sec:integr-cond-LF-LAG-properties}

Let $n$ be a loop of the connectivity graph. From the property Eq.~(\ref{cycle-property-bar-A}), we prove two properties of the good MTLZ families.

{Property (ii)}: {\it for a good family the associated graph $\Gamma$ does not have loops of length $3$.}

\underline{Proof}: Suppose that there exists a length 3 loop $(\alpha, \beta, \mu)$, with $\alpha = \{a, b\}$, $\beta = \{b, c\}$, and $\mu = \{c, a\}$. If the set of forms $\bar A^{\alpha}$, $\bar A^{\beta}$, and $\bar A^{\mu}$ is linearly independent, then so is the set $\bar A^{\alpha} \otimes \bar A^{\alpha}$, $\bar A^{\beta} \otimes \bar A^{\beta}$, and $\bar A^{\mu} \otimes \bar A^{\mu}$, which contradicts the statement
\begin{eqnarray}
&\label{cycle-property-3-loop}\!   s_{ab} \bar A^{ab} \otimes \bar A^{ab} \!+\!s_{bc} \bar  A^{bc} \otimes \bar A^{bc}+ s_{ca}\bar  A^{ca} \otimes \bar  A^{ca}\! = \!0,
\end{eqnarray}
obtained by applying Eq.~(\ref{cycle-property-bar-A}) to the cycle with only three edges. If only two of them, say $A^{ab}$ and $A^{ac}$ are linearly independent, then so is the set, represented by $\bar A^{ab} \otimes \bar A^{ab}$, $\bar A^{ac} \otimes \bar A^{ac}$, and $\bar A^{ab} \otimes \bar A^{ac} + {\bar A^{ac} \otimes \bar A^{ab}}$. Let  $\bar A^{bc} = \lambda_{b} \bar A^{ab} + \lambda_{c} \bar A^{ac}$, for some numbers $\lambda_{b}$ and $\lambda_{c}$. Then Eq.~(\ref{cycle-property-3-loop}) leads to
\begin{align}
\label{cycle-property-3-loop-2}
\nonumber
& \left(s_{ab} +s_{bc}\lambda_{b}^{2}\right) \bar A^{ab}\otimes \bar A^{ab} 
+ \left(s_{ca}  +
 s_{bc}\lambda_{c}^{2} \right) \bar A^{ac} \otimes \bar \bar A^{ac} \\
& + s_{bc} \lambda_{b} \lambda_{c} (\bar  A^{ab} \otimes \bar A^{ac} + {\bar A^{ac} \otimes \bar A^{ab}})  = 0.
\end{align}

Obviously, at least one coefficient in the linear combination in Eq.~(\ref{cycle-property-3-loop-2}) is nonzero, which contradicts the linear independence of the three quadratic forms in Eq.~(\ref{cycle-property-3-loop-2}). So we are left with the only option that any pair of forms among $\bar A^{\alpha}$, $\bar A^{\beta}$, and $\bar A^{\mu}$ is linearly dependent, which contradicts the good family assumption. Therefore we conclude that a length 3 loop does not exist.

{Property (iii)}: let $(\alpha, \nu, \beta, \mu)$ be a loop of the good family graph $\Gamma$ of length $4$.
Then  {\it the vector space spanned by the set $\{\bar A^{\alpha}, \bar A^{\beta}, \bar A^{\mu}, \bar A^{\nu}\}$ has dimension $2$.}

\underline{Proof}:  We denote $\alpha = \{a, b\}$, $\nu = \{b, v\}$, $\beta = \{v, u\}$, and $\mu = \{u, a\}$. Applying Eq.~(\ref{cycle-property-bar-A}) to our cycle, we obtain an analogue of Eq.~(\ref{cycle-property-3-loop})
\begin{align}
\label{cycle-property-4-loop} 
&s_{ab} \bar A^{ab} \otimes \bar A^{ab} +   s_{bv}\bar A^{bv} \otimes \bar A^{bv} \nn\\
&+ s_{vu} \bar A^{uv} \otimes \bar A^{uv} + s_{ua} \bar  A^{au} \otimes \bar A^{au} = 0,
\end{align}
we see that all four forms may not be linearly independent. Suppose now that exactly three, say $\bar A^{ab}$, $\bar A^{au}$, and $\bar A^{bv}$ are linearly independent. Then we have $\bar A^{uv} = \lambda_{a}\bar  A^{au} + \lambda_{b}\bar  A^{bv} + \lambda_{\alpha} \bar A^{ab}$, and Eq.~(\ref{cycle-property-bar-A}) then reads:
\begin{widetext}
\begin{align}
\label{cycle-property-4-loop-2} & \left(s_{ab} + s_{vu}\lambda_{\alpha}^{2} \right)\bar  A^{ab} \otimes\bar  A^{ab} + \left(s_{ua} + s_{vu}\lambda_{a}^{2}\right) \bar A^{ua} \otimes \bar A^{ua} + \left(s_{bv} + s_{vu} \lambda_{b}^{2}\right) \bar A^{vb} \otimes \bar A^{vb} \nonumber \\
&+  s_{vu}\lambda_{\alpha} \lambda_{a} (\bar A^{ab} \otimes \bar A^{au} + {\bar A^{au} \otimes \bar A^{ab}}) + s_{vu}\lambda_{\alpha} \lambda_{b} (\bar A^{ab} \otimes \bar A^{bv}+ {\bar A^{bv} \otimes \bar A^{ab}}) \nonumber \\
&+ s_{vu}\lambda_{a} \lambda_{b} (\bar A^{au} \otimes \bar A^{bv} + {\bar A^{bv} \otimes \bar A^{au}}) = 0.
\end{align}
\end{widetext}
Since at least one of the coefficients in the linear combination of $6$ quadratic forms in the r.h.s. of Eq.~(\ref{cycle-property-4-loop-2}), which are linearly independent by assumption, is nonzero, we obtain a contradiction. Therefore we are left with two options: the vector space spanned on four linear forms $\{\bar A^{\alpha}, \bar A^{\beta}, \bar A^{\mu}, \bar A^{\nu}\}$ has dimension $1$ or $2$. Dimension $1$ contradicts the assumption of a good family, so that the dimension of the aforementioned space is $2$, which completes the proof.


Property (ii) restricts the geometry of the connectivity graph of integrable models to have no loops with only three edges. Property (iii) is important because according to the integrability condition (\ref{constr-3-LA-bar-A}) every node should belong to some 4-loop of the graph.

\end{document}